\documentclass[usenatbib]{mn2e}
\usepackage{epsfig}
\catcode`\@=11
\def\gta{\ifmmode{\,\mathrel{\mathpalette\@versim>\,}}
    \else{$\,\mathrel{\mathpalette\@versim>}\,$}\fi}
\def\lta{\ifmmode{\,\mathrel{\mathpalette\@versim<\,}}
    \else{$\,\mathrel{\mathpalette\@versim<}\,$}\fi}
\def\@versim#1#2{\lower 2.9truept \vbox{\baselineskip 0pt \lineskip
    0.5truept \ialign{$\m@th#1\hfil##\hfil$\crcr#2\crcr\sim\crcr}}}
\catcode`\@=12  
{\newif\ifnotend
\notendtrue
\def\veclist{ABCDEFGHIJKLMNOPQRSTUVWXYZabcdefghijklmnopqrstuvwxyz.}
\def\top#1#2.{#1}
\def\tail#1#2.{#2.}
\loop\expandafter\xdef\csname v\expandafter\top\veclist\endcsname%
{{\noexpand\bf\expandafter\top\veclist}}
\edef\veclist{\expandafter\tail\veclist}
\if\veclist.\notendfalse\fi\ifnotend\repeat}
\def\fracj#1#2{{\textstyle{#1\over#2}}}
\def\Cov{{\rm Cov}}\def\Var{{\rm Var}}

\def\vpbar{\overline{v}_\phi}
\def\Ug{U_{\rm g}}\def\Vg{V_{\rm g}}\def\Wg{W_{\rm g}}

\def\vmu{\bf\mu}\def\ex#1{\langle#1\rangle}

\def\kms{\,{\rm km}\,{\rm s}^{-1}}

\def\kpc{\,{\rm kpc}}

\def\e{{\rm e}}
\def\d{{\rm d}}

\def\figref#1{Fig.~\ref{#1}}
\def\vpa{v_{\parallel}}

\def\ovpas{\overline{\vpa^2}}

\def\ovs{\overline{v^2}}
\def\ovfs{\overline{F^2}}

\def\ovcv0{\overline{V_0}}

\newcommand{\beq}{\begin{equation}}
\newcommand{\eeq}{\end{equation}}

\title[The detection and treatment of distance errors]
{The detection and treatment of distance errors in kinematic analyses of stars}

\author[Sch\"onrich, Binney \& Asplund]{Ralph Sch\"onrich$^1$\thanks{E-mail:rasch@mpa-garching.mpg.de}, James Binney$^2$ and Martin Asplund$^1$\\
 $^{1}$ Max Planck Institute for Astrophysics, Karl-Schwarzschild-Str. 1, D-85741 Garching\\ 
$^{2}$ University of Oxford, Rudolf-Peierls Centre for Theoretical Physics, Keble Road, Oxford OX1 3NP\\}

\begin{document}

\date{October 28th, 2011}

\pagerange{\pageref{firstpage}--\pageref{lastpage}} \pubyear{2011}

\maketitle

\label{firstpage}

\begin{abstract}
 We present a new method for detecting and correcting systematic errors in
the distances to stars when both proper motions and line-of-sight velocities
are available. The method, which is applicable for samples of $200$ or more
stars that have a significant extension on the sky, exploits correlations
between the measured $U,V$ and $W$ velocity components that are introduced by
distance errors. We deliver a formalism to describe and interpret the
specific imprints of distance errors including spurious velocity correlations
and shifts of mean motion in a sample.  We take into account correlations
introduced by measurement errors, Galactic rotation and changes in the
orientation of the velocity ellipsoid with position in the Galaxy. Tests on
pseudodata show that the method is more robust and sensitive than traditional
approaches to this problem. We investigate approaches to characterising the
probability distribution of distance errors, in addition to the mean distance
error, which is the main theme of the paper. Stars with the most
overestimated distances bias our estimate of the overall distance scale,
leading to the corrected distances being slightly too small. We give a
formula that can be used to correct for this effect.  We apply the method to
samples of stars from the SEGUE survey, exploring optimal gravity cuts,
sample contamination, and correcting the used distance relations.
\end{abstract}

\newcounter{mytempeqncnt}
\section{Introduction}

Studies of stellar kinematics in the Milky Way are of enormous
importance as they hold the key both to measuring the gravitational field
of the Galaxy and to unravelling the Galaxy's history and manner of
formation. Consequently considerable resources have been, and are being,
devoted to measuring the velocities of stars. 

Two different techniques have to be used to measure the three components of
velocity with respect to the Sun: the component $v_\parallel$ along the line
of sight to the star is measured spectroscopically, while the component
$\vv_\perp$ transverse to the line of sight is determined by combining the
measured proper motion $\vmu$ with an estimate of the distance $s$ to the
star. Over the next decade enormous numbers of distances will be obtained
from parallaxes measured by the Gaia satellite, but currently the great
majority of distance estimates have been obtained by comparing an estimate of
the star's absolute magnitude with its apparent magnitude. This process is
liable to systematic error in several ways. Giants can be mistaken for
subgiants or even dwarfs of the same colour (or vice versa) and assumed
ages severely influence the adopted luminosities in the turn-off region
even for well classified stars. Also, the adopted metallicities
may be biased as discussed by \cite{Lee08a} and as demonstrated by the shifts
in metallicity scale between \cite{Nordstrom04}, \cite{HolmbergNA} and
\cite{CSA11}. An erroneous metallicity will lead to the wrong isochrone being
used to infer the luminosity, and an erroneous luminosity and distance will
follow. Further problems are that synthetic colours can be wrong \citep[cf.
the discussion in][]{Percival09} and that stellar-evolution models can
predict different luminosities for given metallicity and effective
temperatures; there is evidence that they make the main sequences of
metal-poor objects too faint \citep[the ``helium problem'', e.g.  discussed
in][]{Casagrande07}. Finally, erroneous extinctions may be adopted.
Since the problems just enumerated can readily accumulate to systematic
distance biases in excess of $20$ per cent, some way of independently
calibrating the distance scale is invaluable. 

Here we present a method for calibrating distances that exploits
correlations between the measured $U,V,W$ components of velocity that are
introduced by systematic distance errors, and is applicable to any survey
that provides proper motions and line-of-sight velocities over a wide area of
the sky.

The idea that the typical distance to objects in a sample can be constrained
by proper motions is well known in astronomy -- for a useful recent review
see \cite{Popowski98}. The method of secular parallaxes determines the mean
parallax of a population by combining proper motions and the known mean
motion of the population with respect to the Sun
\citep[e.g.][\S2.2.3]{Trumpler62,BinneyMerrifield}, while the method of
statistical parallaxes estimates the mean parallax by combining proper
motions with line-of-sight velocities
\citep[e.g.][\S2.2.4]{BinneyMerrifield}. Our method has points in common with
both the above methods in that it hinges on comparing proper motions with
line-of-sight velocities but also exploits the mean motion of the stars with
respect to the Sun. It is much less vulnerable than classical methods to
questionable assumptions regarding the shape of the velocity ellipsoid and/or
the nature of mean velocity field \citep[see esp. the discussion
in][]{Trumpler62}. By examining the way correlations between components of
space velocity vary with position on the sky, we dispense with the need for
prior knowledge of the mean velocity field.  All we require is knowledge of
the formal errors of the observables and, if the sample is sufficiently
non-local, reasonable assumptions about the orientation of the velocity
ellipsoid at relevant points in the Galaxy.


Section \ref{sec:af} lays out the basic theory for the case in which
distances are all in error by a common factor. Section \ref{sec:scatter}
extends the theory to the realistic case in which distance errors contain a
random component. Section \ref{sec:implement} applies the method to data from
the Sloan surveys. Section \ref{sec:conclude} sums up.

\section{The mean distance error}\label{sec:af}

We are concerned with the case in which
calibration errors in the distance scale cause  all
distances have a fractional error $f$, so the assumed distance $s'$ to a star
is related to the true distance $s$ by
 \begin{equation}
s'=(1+f)s.
\end{equation}
 Consequently the assumed tangential velocity $\vv_\perp'$ is related to the
true tangential velocity $\vv_\perp$ by
\begin{equation}
\vv_\perp' = (1+f)\vv_\perp.
\end{equation}
 The velocity component $\vpa$ along the line of sight is of course
unaffected by distance errors. 

From $\vpa$ and the proper motions $(\mu_b=\dot b,\mu_l=\cos b\,\dot l)$
parallel to each Galactic coordinate we infer the velocity components
$(U,V,W)$ in the Cartesian coordinate system in which the Sun is at rest at
the origin. In this system the $U$ axis points to the Galactic centre, the
$V$ axis points in the direction of Galactic rotation, and the $W$ axis
points to the north Galactic pole. The relevant transformation is 
 \begin{equation}
\pmatrix{ U_0 \cr V_0 \cr W_0} = \vM
\pmatrix{s\mu_b \cr s\mu_l \cr v_{\parallel}},
\end{equation}
 where the orthogonal matrix
\begin{equation}
\vM\equiv\pmatrix{
-\sin b \cos l  & -\sin l  & \cos b \cos l  \cr
-\sin b \sin l  & \cos l  & \cos b \sin l  \cr
\cos b  & 0 & \sin b }.
\end{equation}
 The velocity components inferred from distances that have fractional error
$f$ are
 \begin{equation}
\pmatrix{ U \cr V \cr W} =
 \vM (\vI+ f\vP) \pmatrix{s\mu_b \cr s\mu_l \cr v_{\parallel}},
\end{equation}
 where $\vI$ is the identity matrix and 
\begin{equation}
\vP\equiv\pmatrix{1&0&0\cr0&1&0\cr0&0&0}.
\end{equation}
 Hence the  true and measured Galactocentric components of
 velocity are related by
\begin{equation}\label{eq:UfromU0}
\pmatrix{ U \cr V \cr W} =
\vM(\vI+ f\vP)\vM^T\pmatrix{ U_0 \cr V_0 \cr W_0}
=(\vI+f\vT)\pmatrix{ U_0 \cr V_0 \cr W_0},
\end{equation}
where
\begin{equation}
\vT\equiv\vM\vP\vM^T.
\end{equation}
 Table \ref{tab:mat} gives an explicit expression for $\vT$, which has
direction-dependent off-diagonal elements.  Consequently, when $f\ne0$ the
inferred value of $W$ has linear dependencies on $U_0$ and $V_0$ with
coefficients that are known functions of Galactic position times $f$. By detecting
these patterns of bias, we can measure the amount $f$ by which distances have
been overestimated. 

The phenomenon we exploit can be understood by an example. Consider a star at
a Galactic longitude $l=0$ and latitude $b=45^\circ$. Suppose the star's only
non-zero component of velocity (in the Sun's rest frame) is $U_0>0$. This motion generates both a
proper motion $\mu_b<0$ and a line-of-sight velocity $\vpa$ away from us. If we
overestimate the star's distance, the tangential velocity, which lies in the
$(U,W)$ plane, will be overestimated, and we will infer a negative value for
$W$ instead of zero. By the same token, a star with overestimated distance that had $U_0<0$ would
have $W>0$. In the southern Galactic hemisphere signs reverse and a star with
overestimated distance at $b=-45^\circ$ with $U_0>0$ will be wrongly assigned a positive value of $W$.  Hence a systematic tendency to misjudge
distances can be detected by looking for correlations between velocity
components that vary over the sky in given ways.

\begin{table*}
\caption{Explicit expression for the matrix $\vT$ through which distance
errors introduce correlations between the apparent components of velocity,
and an expression for the relation between the errors in $(U,V,W)$ and in
$(\mu,\vpa)$.}
\begin{tabular}{c}
$
\vT = \vM\vP\vM^T = \pmatrix{
1-\cos^2b\cos^2l  & -\frac12\cos^2b\sin 2l
& -\frac12\sin 2b \cos l  \cr
-\frac12\cos^2 b \sin 2l  &1-\cos^2 b \sin^2 l
& -\frac12\sin 2b \sin l  \cr
-\frac12\sin 2b \cos l  & -\frac12\sin 2b \sin l  & \cos^2b }\nonumber
$\\ \\$
\pmatrix{e_U \cr e_V \cr e_W} =\vM(\vI+ f\vP)\pmatrix{s\epsilon_b\cr
s\epsilon_l\cr \epsilon_\parallel}=  \pmatrix{
-\sin b \cos l (1+f) s \epsilon_b -\sin l (1+f) s \epsilon_l + \cos b \cos l \epsilon_\parallel \cr
-\sin b \sin l (1+f) s \epsilon_b + \cos l (1+f) s \epsilon_l + \cos b \sin l \epsilon_\parallel \cr
\cos b (1+f) s \epsilon_b + \sin b  \epsilon_\parallel  
}\nonumber
$\\
\end{tabular}\label{tab:mat}
\end{table*}

The Sun moves in the direction of Galactic rotation faster than the circular
speed and all Galactic components are subject to at least some asymmetric
drift, so $\ex{V_0}<0$ for most groups of stars, especially halo stars.
Consequently, the clearest signals of an erroneous distance scale are usually
correlations between the measured values of $U$ and $V$ and between $W$ and
$V$. 

\subsection{A naive Approach}\label{sec:naive}

Equations (\ref{eq:UfromU0}) yield
 \begin{eqnarray}\label{eq:basicUVW}
U&=&(1 + fT_{UU})\,U_0 +fT_{UV}V_0+fT_{UW}W_0\nonumber\\
V&=&(1+fT_{VV})V_0+fT_{VU}U_0+fT_{VW}W_0\\
W&=&(1+fT_{WW})\,W_0+fT_{WU}U_0+T_{WV}V_0.\nonumber
\end{eqnarray}
 Suppose for each hemisphere $b>0$ and $b<0$ we bin
stars in $l$ and in $V\simeq V_0$. Then for each bin we could average the
first and third equations, obtaining  for each bin two equations
 \begin{eqnarray}\label{eq:avset}
\ex{U}&=&\ex{(1+fT_{UU})\,U_0}+f\ex{T_{UV}V_0}+f\ex{T_{UW}W_0}\nonumber\\
\ex{W}&=&\ex{(1+fT_{WW})\,W_0}+f\ex{T_{WU}U_0}+f\ex{T_{WV}V_0}.
\end{eqnarray}
 We expect the population of stars under study, taken as a whole, to be
moving neither radially nor vertically, so at any $(b,l)$ the mean values of
$U_0$ and $W_0$ should be the reflex of the solar motion,
$(U_\odot,W_\odot)$. With this assumption in the $U$ equation we may set
 \begin{eqnarray}\label{eq:keya}
\ex{(1+fT_{UU})U_0}&=&-(1+f\ex{T_{UU}})U_\odot\nonumber\\
\ex{T_{UW}W_0}&=&-\ex{T_{UW}}W_\odot,
\end{eqnarray}
 and similar relations can be used in the $W$ equation. Finally we make $f$ the
only unknown in equations (\ref{eq:avset}) by assuming, in a first
approximation, that $V_0=V$. On account of Poisson noise, the sample values
of quantities such as $\ex{T_{UW}W_0}$ will differ from our adopted
value, $-\ex{T_{UW}}W_\odot$, so the equations will not be exactly
satisfied,  but we can seek the values of $f$ that minimise the
quantities
 \begin{eqnarray}
S'_U&\equiv&\sum_{\rm
bins}\bigl[\ex{U}+(1+f\ex{T_{UU}})U_\odot\nonumber\\
&&\qquad-f\ex{T_{UV}\,V}+f\ex{T_{UW}}W_\odot\bigr]^2\\
S'_W&\equiv&\sum_{\rm bins}\bigl[\ex{W}+(1+f\ex{T_{WW}})W_\odot\nonumber\\
&&\qquad+f\ex{T_{WU}}U_\odot
-f\ex{T_{WV}V}
\bigr]^2\nonumber.
\end{eqnarray}
 After determining the optimum value of $f$,
this value can be used to correct the distances and the velocities derived
from them,
and a new value of $f$ is then obtained, enabling the distances to be corrected a
second time, and so on until convergence has been reached.

The scheme just described is straightforward conceptually and does work, but
suffers a significant loss of information from the need to bin the data and to replace
the measured values of $U$ and $W$ by $-U_\odot$ and $-W_\odot$. Therefore
the results shown in this paper are obtained by a different scheme that is
described in the next subsection.

\subsection{A more effective approach}\label{sec:sophisticated}

Our method is based on the principle that the true value of $U$ or $W$
can be decomposed into a mean velocity field of known form with components
$\overline{U}$ and $\overline{W}$, plus a random variable $\delta U$ or
$\delta W$ that has zero mean, so $U_0={\overline{U}}+\delta U$, etc. In this
subsection we make the assumption that the mean velocity field may be
approximated by the reflex of the solar motion, so $\overline{U} = -U_\odot$
etc. In Section \ref{sec:rotation} we will lift this restriction to allow for
Galactic rotation. Also we argue that in the second or third terms on the
right of equations (\ref{eq:basicUVW}) we may replace $V_0$ by $V$ on the
ground that the inferred value is close to the true value and the presence in
these terms of an explicit factor $f$ implies that the error made by
replacing $V_0$ by $V$ is O($f^2$). The same argument enables us to replace
$U_0$ by $U$ and $W_0$ by $W$ in these terms.  With these replacements the
first and third of equations (\ref{eq:basicUVW}) become
 \begin{eqnarray}\label{eq:otherUW}
U&=&-U_\odot + fx+(1+fT_{UU})\delta U\nonumber\\
W&=&-W_\odot+fy+(1+fT_{WW})\delta W,
\end{eqnarray}
 where 
\begin{eqnarray}\label{eq:defsxy}
x&\equiv&-T_{UU}U_\odot+T_{UV}V +T_{UW}W\nonumber\\
y&\equiv&-T_{WW}W_\odot+T_{WU}U+T_{WV}V.
\end{eqnarray}
 We now eliminate reliance on prior knowledge of the solar motion by
subtracting from each of equations (\ref{eq:otherUW}) its expectation value,
and have
\begin{eqnarray}
U-\ex{U}&=&f(x-\ex{x})+(1+fT_{UU})\delta U\nonumber\\
W-\ex{W}&=&f(y-\ex{y})+(1+fT_{WW})\delta W,
\end{eqnarray}
 We determine the optimum value of $f$ by forming the sample
sums\footnote{The accuracy with which $f$ is determined can be slightly
increased by weighting each term in the sums in eqs.(\ref{eq:wsums}) by the
inverse of the expected standard deviation of the noise term or by using
Huber-White standard errors \citep{White80}.}
 \begin{eqnarray}\label{eq:wsums}
\sum_i[U_i-\ex{U}-f(x_i-\ex{x})]x_i&=&\sum_i(1+fT_{UUi})\delta U_ix_i
\\
\sum_i[W_i-\ex{W}-f(y_i-\ex{y})]y_i&=&\sum_i(1+fT_{WWi})\delta W_iy_i.\nonumber
\end{eqnarray} 
 The right side of the first equation would vanish if $\delta U$ were
uncorrelated with $x$ but it {\it is\/} correlated because $x$ depends on $V$
and $W$, which in turn depend on $U_0=-U_\odot+\delta U$. In fact one easily
shows that
 \begin{equation}
\ex{(1+fT_{UU})\delta Ux}=f\ex{(1+fT_{UU})(T_{UV}^2+T_{UW}^2)\delta U^2}.
\end{equation}
 Since we are working to O($f$) only, we neglect the second term in the first
bracket on the right and use the resulting expression in equation
(\ref{eq:wsums}) to solve for $f$. We find
 \begin{equation}\label{eq:ffromU}
f={\Cov(U,x)\over\Var(x)+\ex{T_{UV}^2+T_{UW}^2}\sigma_U^2},
\end{equation}
 where $\sigma_U^2\equiv\ex{\delta U^2}$ and we have identified sample means
with expectation values. Analogously, from the second of equations
(\ref{eq:wsums}) we have
 \begin{equation}\label{eq:ffromW}
f={\Cov(W,y)\over\Var(y)+\ex{T_{WV}^2+T_{WU}^2}\sigma_W^2}.
\end{equation}
 As with the naive scheme, we proceed iteratively, successively correcting
the distances according to the value of $f$ yielded by the current distances,
until $f$ becomes negligible.  The precise values of the denominators in our
expressions for $f$ are not important because we rescale distances until
$f\propto\Cov(U,x)=0$ or $f\propto\Cov(W,y)=0$. This circumstance is
fortunate as only an approximate values of $\sigma_U$ and $\sigma_W$ may be
available. Note that equations (\ref{eq:ffromU}) and (\ref{eq:ffromW}) make
no reference to the solar motion so that in contrast to the secular parallax
they require only information about the shape, but not the average value of the mean
velocity field.

In the following, when using equation (\ref{eq:ffromU}) we shall call $U$ the
``target variable'' and $x$ the ``explaining variable'', while when we use
equation (\ref{eq:ffromW}), $W$ will be the target variable and $y$ the
explaining variable.

The reader may wonder why we do not obtain a third estimate of $f$ from the
$V$ equation of the set (\ref{eq:basicUVW}). The problem is that we cannot
write $V_0=-V_\odot+\delta V$ by analogy with our treatment of $U$ and $W$,
because most stellar groups have mean azimuthal velocities smaller than that
of the Sun, and in fact the mean azimuthal velocity of a group will vary with
location.

The quantities $\sum_iU_ix_i$ and $\sum_iW_iy_i$ implicit in the right sides
of equations (\ref{eq:ffromU}) and (\ref{eq:ffromW}) contain cross-terms such
as $\sum_iU_iV_i$ and $\sum_iU_iW_i$.  As explained above, usually the $V$
cross-terms contain the largest amount of information regarding $f$, except
when $V$ lies near zero, when the $W$ cross-terms provide the strongest
constraints on $f$.  $W$ is the target velocity of choice both because
it has the lowest velocity dispersion, and because it is least affected by
streaming motions, which are largely confined to the $UV$ plane
\citep{Dehnen98}.

\begin{figure}
\epsfig{file=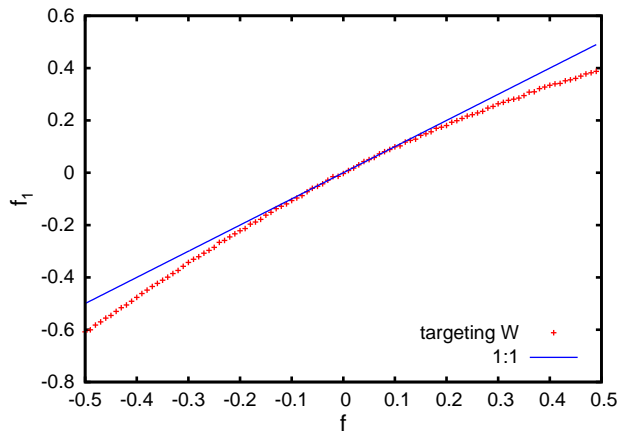,angle=-90,width=\hsize}
 \caption{Value of the fractional distance error $f_1$ from equation
(\ref{eq:ffromW}) versus value of $f$ bias in a mock
sample of $450\,000$ disc and $50\,000$ halo stars. The blue line has unit
slope.}\label{fig:simplef}
\end{figure}

\begin{table}
\caption{Parameters of the mock disc and halo samples used in tests. All
velocities are in $\!\kms$}
\label{tab:model}
\begin{tabular}{lcccc}
Component&$\sigma_U$&$\sigma_V$&$\sigma_W$&$\overline{v}_\phi$\\
Disc&55&45&35&180\\
Halo&150&75&75&0\\
\end{tabular}
\end{table}

\subsection{Tests}\label{sec:tests}

In this section we test the effectiveness of the scheme derived in the
last subsection by deriving pseudo-data from a model Galaxy, and analysing
them in the presence of systematic distance errors. We have conducted such
tests using a model obtained by adding gas and star formation to a halo
formed in simulations of the cosmological clustering of collisionless
particles. The results of these tests were entirely satisfactory, but we do
not report them here for two reasons: (a) considerable space would be
required to describe the Galaxy model with sufficient precision and the model
is in any case not entirely realistic, and (b) the model provides a rather
limited number of particles in the vicinity of the Sun, so the statistical
precision of the tests is inferior to that of the tests we will present.
These use data obtained from a Galaxy model that is highly idealised, but
which has the flexibility to produce data that include or exclude whatever
features in the data might affect the performance of our method.

Our idealised Galaxy model has a non-rotating halo and a rotating disc. The velocity
ellipsoids of both components are triaxial Gaussians: Table~\ref{tab:model}
gives the values of the dispersions. The mean rotation velocity of the disc
is taken to be $180\kms$ and the circular speed is $220 \kms$.  The sampled
stars are distributed uniformly in distance between $0.5 \kpc$ and $4\kpc$,
and uniformly in Galactic longitude and latitude, which gives the sample a
strong poleward bias that resembles the bias encountered in real samples
better than an isotropically distributed sample would. The solar motion is in
addition offset by the local standard of rest velocity vector as determined
by \cite{SBD}.

The crosses in \figref{fig:simplef} show the value of $f$ recovered from
equation (\ref{eq:ffromW}) on the first iteration, $f_1$, versus the preset fractional distance
overestimate $f$ applied to the sample. Each cross shows an independent realisation of the
pseudodata, which contained $450\,000$ disc and $50\,000$ halo stars.  The crosses fall
on a curve that passes through $(0,0)$ as we would hope. The straight line
through the origin with unit slope is also plotted and we see that for
$|f|\lta0.2$ the slope of the curve is close to unity, so convergence of the
iterative procedure is rapid. However, the key point is that the
curve passes through the origin and has no point of inflection. So long
as these conditions are satisfied, the iterative scheme will converge on the
correct distance scale regardless of the slope or curvature of the curve.

\begin{figure}
\epsfig{file=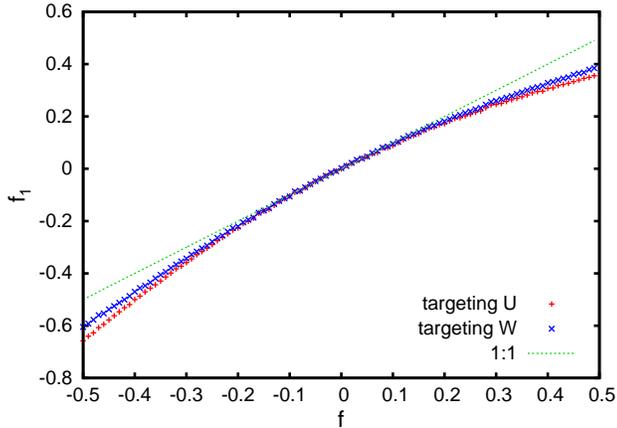,angle=-90,width=\hsize}
 \caption{Value of the fractional distance error $f$ from equation
(\ref{eq:ffromU}) or (\ref{eq:ffromW}) versus the input value of $f$ for
samples of $500\,000$ stars with an isotropic velocity distribution and no
solar motion. The line of unit slope is also shown.}\label{fig:isotrop}
\end{figure}

\figref{fig:isotrop} demonstrates that the method works well even in the
absence of solar motion by showing results analogous to those of
\figref{fig:simplef} for a sample of stars that has no net motion with
respect to the Sun and an isotropic velocity distribution around the solar
motion, i.e. without any systematic offset. The minor difference between the
estimators on $U$ and $W$ derives from second-order effects in $f$ by the polewards bias in the sample geometry. Note that a simple linear regression of $W$ on $y$ from equation
(\ref{eq:otherUW}) would give the right zero point and so finally an unbiased
distance estimate, but due to the lack of correction factors to the
denominator $\Var(x)$ would give a slope that is a factor $2$ too large and
hence a bad convergence behaviour in the iteration.

\subsection{Impact of random errors}\label{sec:errors}

We now consider the impact on our technique of random measurement errors. If
we measured $U$, $V$ and $W$ directly, random errors would have no impact
because they would simply inflate the scatter in these variables that is
inherent in stars having random velocities. Unfortunately, we do not measure
$U,V,W$ directly but calculate them from the measured values of
$\mu_b,\,\mu_l$ and $\vpa$. Consequently the error in say $\mu_l$ introduces
correlated errors into both $U$ and $V$. Since our technique consists
precisely in attributing correlations between $U$ and $x$ (which depends on
$V$) to a non-zero value of $f$, we must consider the contribution of the
errors to the correlations between $U$ and $x$ or $W$ and $y$ if we are to
estimate $f$ correctly.

Our key assumption is that the errors in $\mu_b,\,\mu_l$ and $\vpa$ are
statistically independent, have vanishing mean, and have finite and
approximately known variances. Let
$\epsilon_b,\epsilon_l,\epsilon_\parallel$ be the errors in the proper
motions and line-of-sight velocity. Then the random errors in $U,V,W$ are
 \begin{equation}\label{eq:errorUVW}
\pmatrix{ e_U \cr e_V \cr e_W} =
\vM(\vI+ f\vP)\pmatrix{s\epsilon_b\cr s\epsilon_l\cr \epsilon_\parallel}.
\end{equation}
 Table~\ref{tab:mat} gives an explicit expression for the right side of this
equation.

Consider now the correlation, between a target variable, say $W$, and
the explaining variable $y$. Let $W = W^\prime+ e_W$ and
$y = y^\prime + e_y$, where the primed variables are the components without
the error and $e_W,\, e_y$ are their errors derived from equation
(\ref{eq:errorUVW})
 \begin{equation}
\ex{Wy}=\ex{W^\prime y^\prime} + \ex{e_We_y} + \ex{W^\prime e_y} + \ex{e_W
y^\prime}.
\end{equation}
Given that the errors are unbiased, the correlations such as $\ex{W'e_y}$
between the true velocities and the errors vanish. Consequently the changes in
$f$ that the errors introduce through equation
(\ref{eq:ffromW}) is
\begin{eqnarray}
e_f&=&{\ex{e_We_y}\over\ex{y^2}+\ex{T_{WU}^2+T_{WV}^2}\sigma_W^2}\nonumber\\
&&\qquad-f{\ex{e_y^2}\over\ex{y^2}+\ex{T_{WU}^2+T_{WV}^2}\sigma_W^2}.
\end{eqnarray}
 The second term on the right side is smaller than the
first, and as we iterate towards $f=0$ it vanishes altogether. Hence we
neglect it. With this term neglected, we can obtain the error-corrected value
of $f$ simply by subtracting $\ex{e_We_y}$ from the measured value of
$\ex{Wy}$ before inserting its value into equation (\ref{eq:ffromW}). 

\begin{figure*}
\epsfig{file=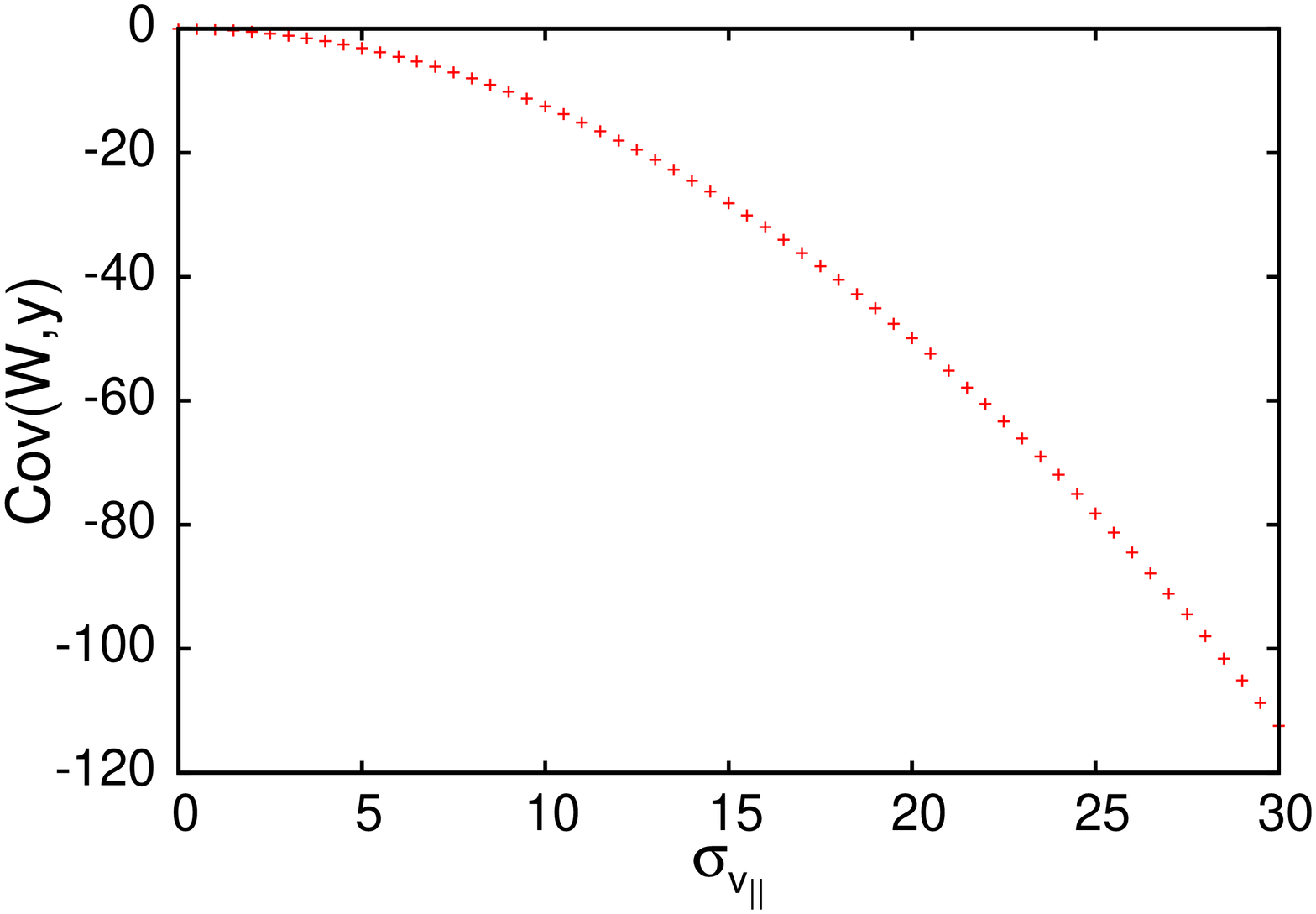,angle=-90,width=0.32\hsize}
\epsfig{file=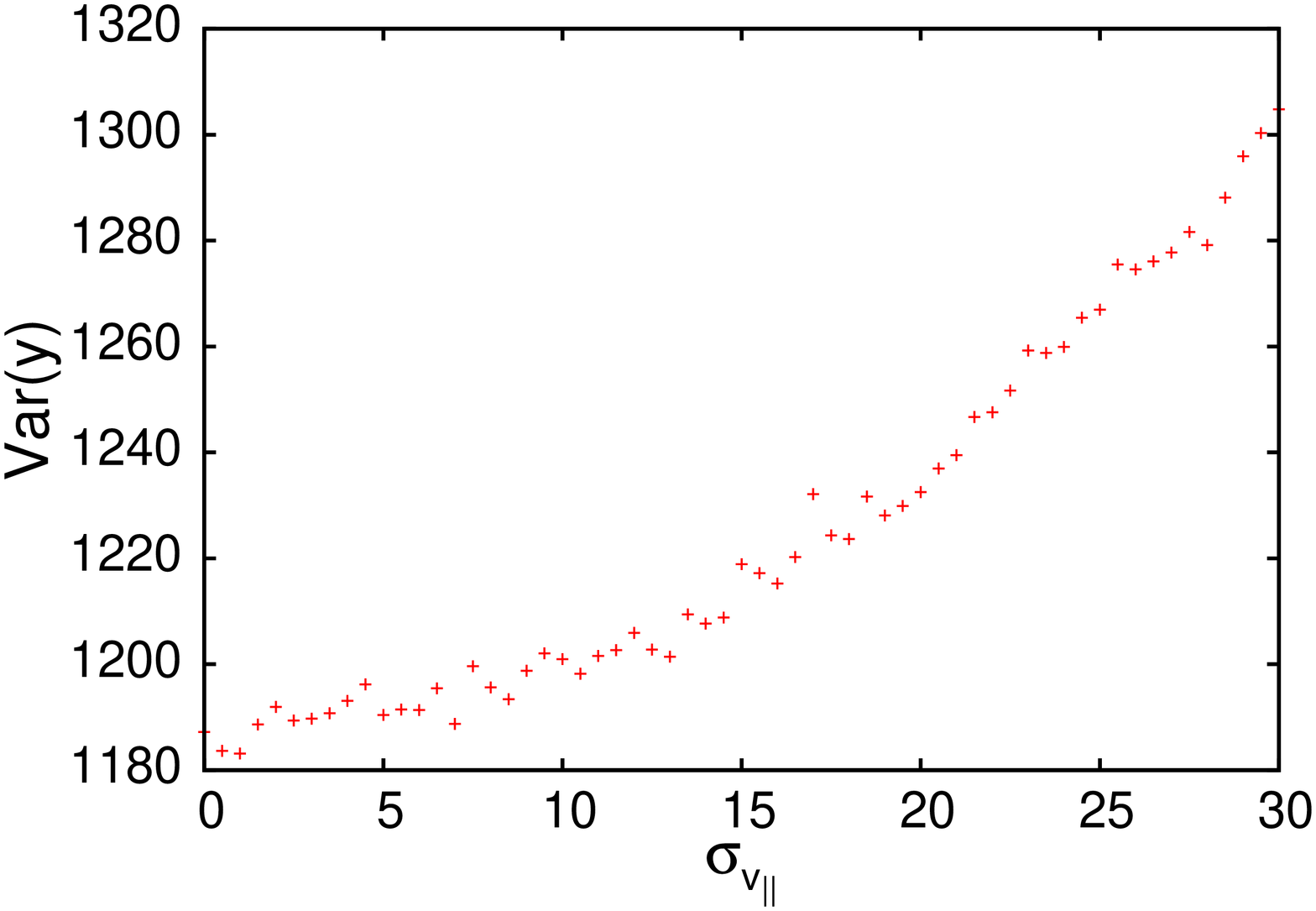, angle=-90,width=0.32\hsize}
\epsfig{file=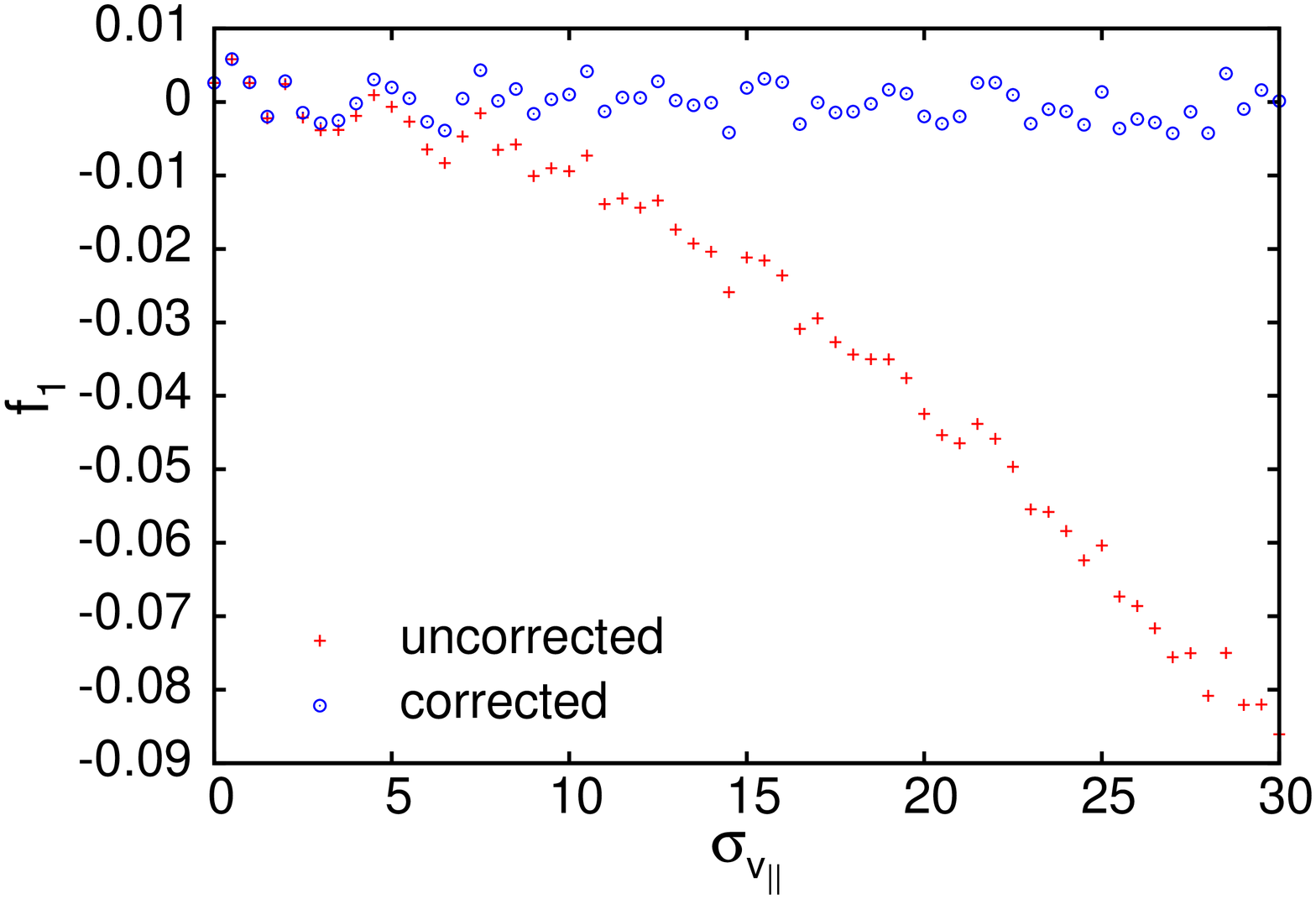,angle=-90,width=0.32\hsize}
\epsfig{file=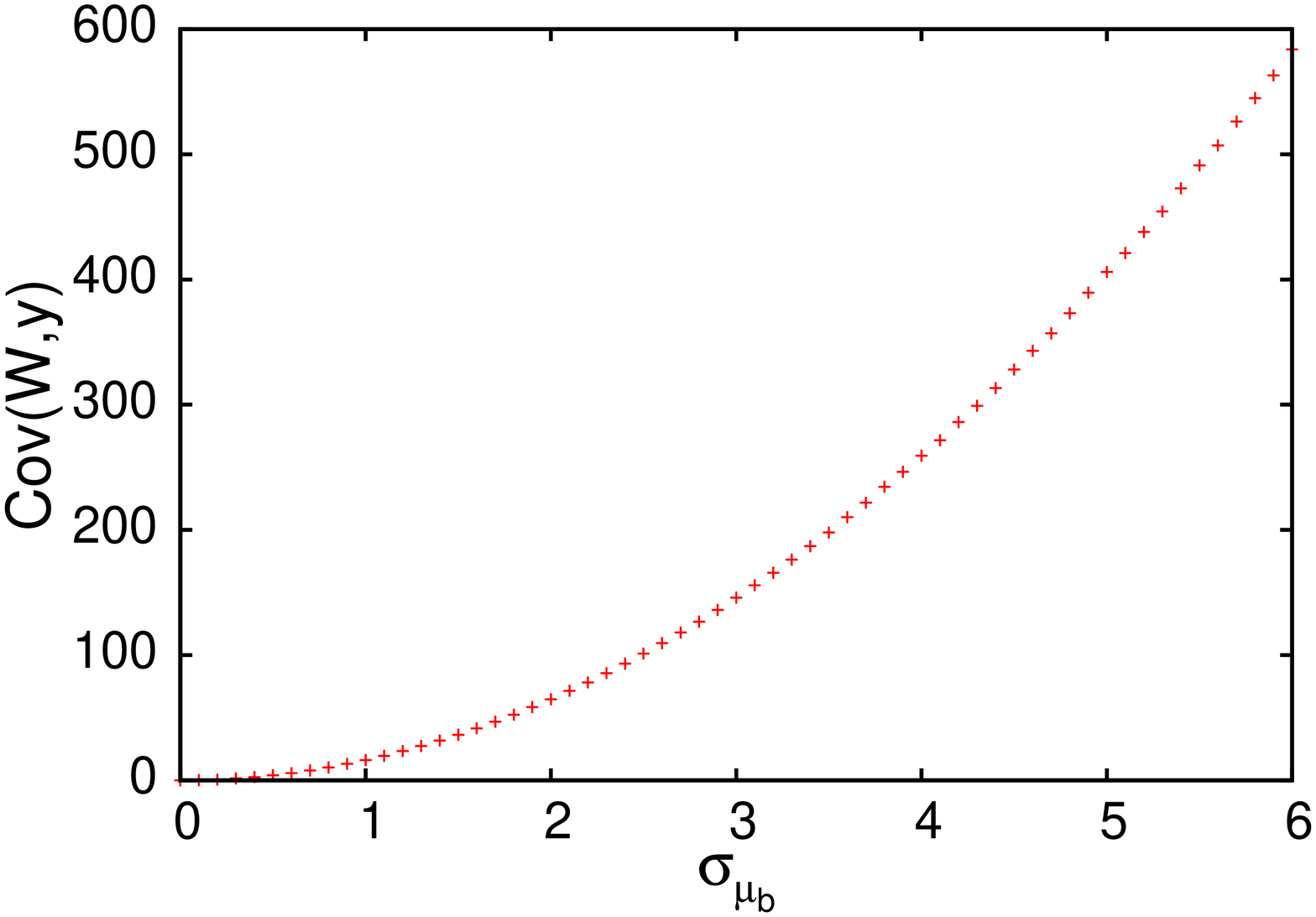, angle=-90,width=0.32\hsize}
\epsfig{file=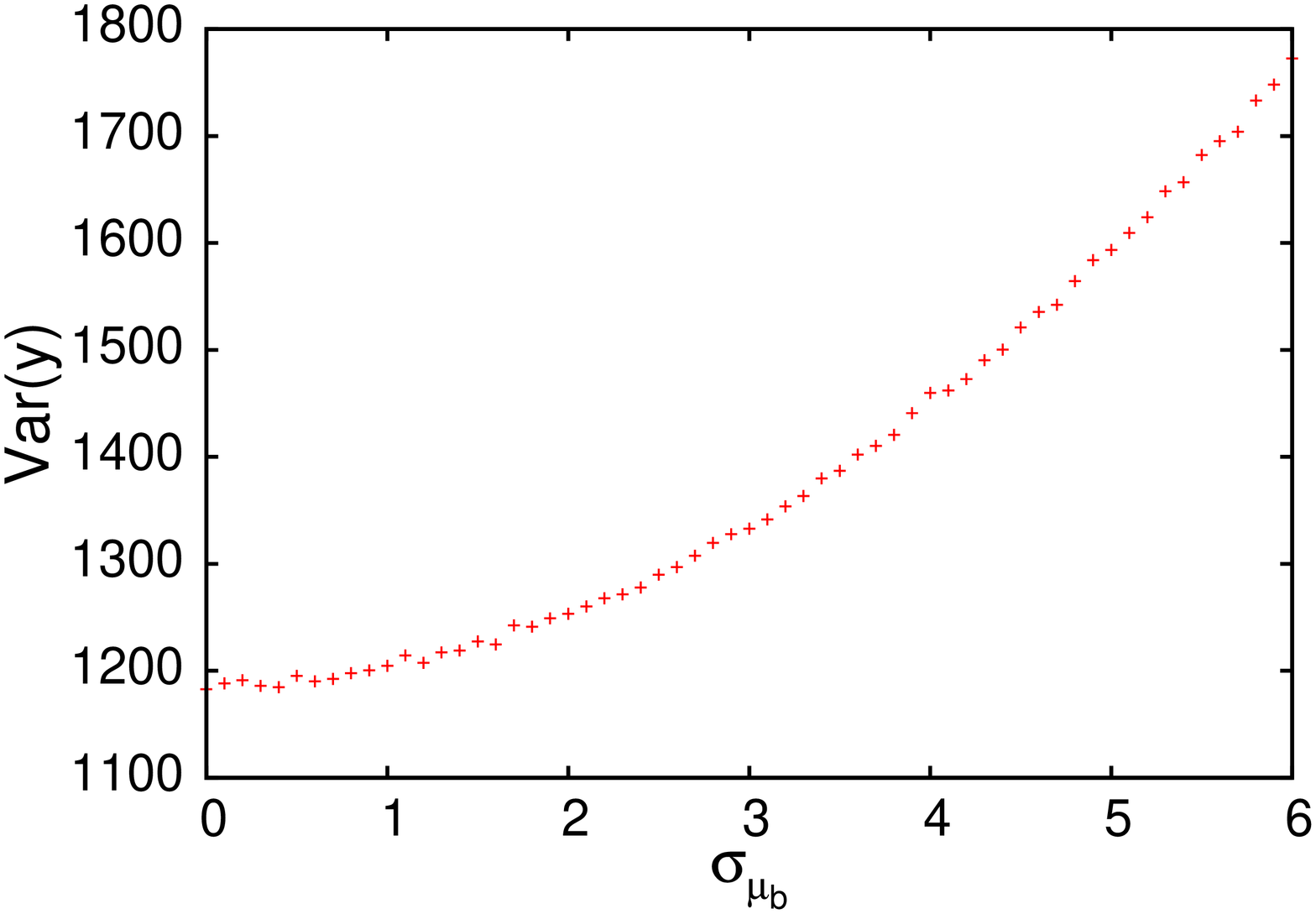,angle=-90,width=0.32\hsize}
\epsfig{file=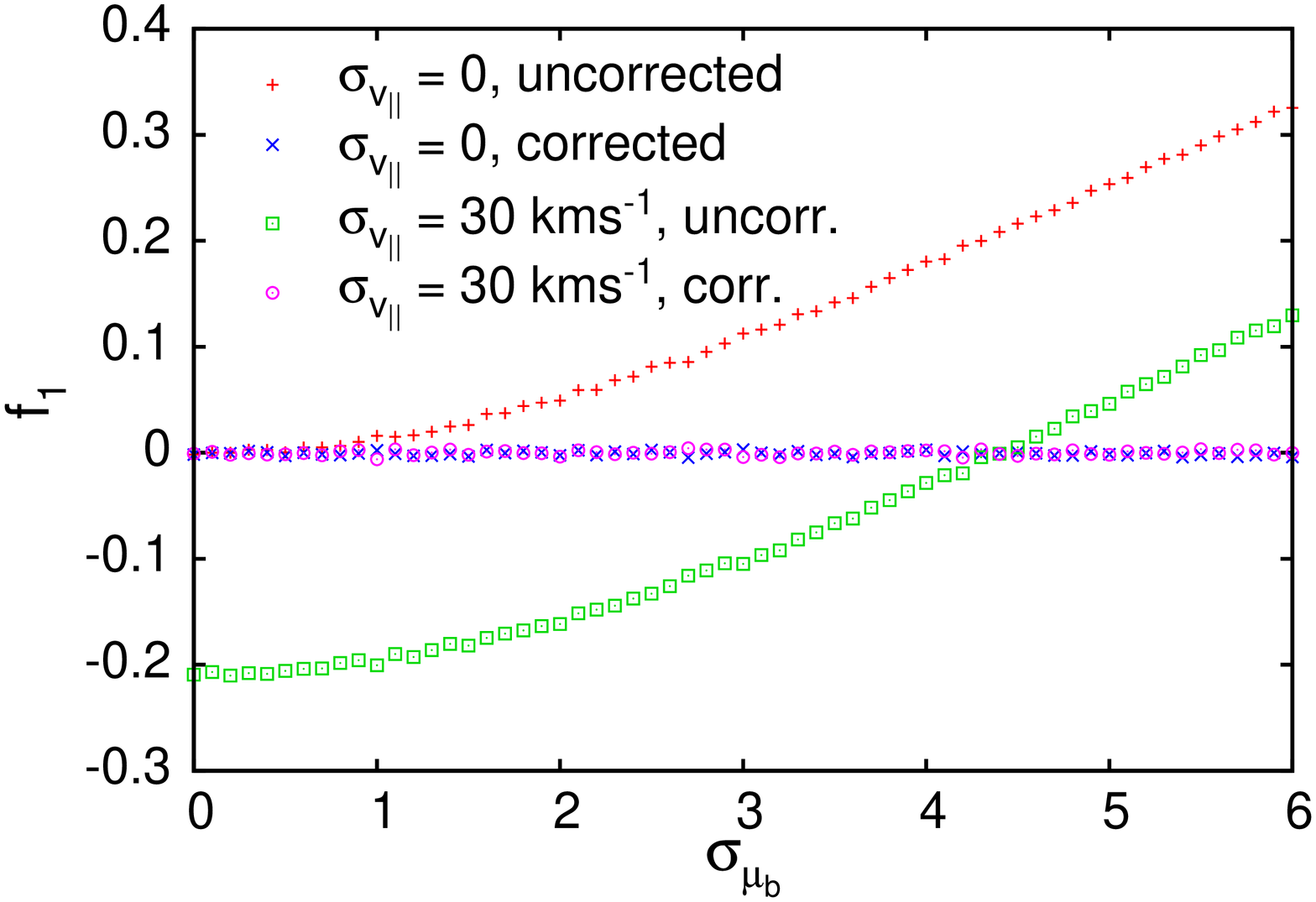, angle=-90,width=0.32\hsize}
 \caption{Tests of the effects of random measurement errors in a mock sample
of $500\,000$ stars, among which are $50\,000$ halo stars.  The left and
centre panels show how the values of $\Cov(W,y)$ and $\Var(y)$ are affected by
measurement errors, while the right panels show the values of $f_1$ using
equation(\ref{eq:ffromW}) with and without the correction terms for these
samples. In the upper row the proper motions are error free and the
horizontal axis gives the error in $\vpa$, while in the lower panel $\vpa$ is
error free (red and green crosses) and the horizontal axis gives the error in
proper motions. In the bottom right panel we added the case of a fixed radial
velocity error of $30 \kms$ (green squares and purple circles) to demonstrate
the simple superposition of the error correlation terms on the covariance.
}\label{fig:errdis}
\end{figure*}

We now calculate the error-error correlations. 
We have from equations (\ref{eq:defsxy})
\begin{eqnarray}
\ex{e_Ue_x}&=&T_{UV}\ex{e_Ue_V} +T_{UW}\ex{e_Ue_W}\nonumber\\
\ex{e_We_y}&=&T_{WU}\ex{e_We_U}+T_{WV}\ex{e_We_V}.
\end{eqnarray}
 When we use Table \ref{tab:mat} to express the errors in terms of the
(uncorrelated) errors
in the observables, we find
\begin{eqnarray}
\ex{e_Ue_V}&=&\fracj12\sin2l[(1+f)^2 s^2(\sin^2b\epsilon_b^2
-\epsilon_l^2)+\cos^2b\epsilon_\parallel^2]\nonumber\\
\ex{e_Ue_W}&=&
-\fracj12\sin2b\cos l[(1+f)^2s^2\epsilon_b^2 - \epsilon_\parallel^2]\\
\ex{e_We_V}&=&
-\fracj12\sin2b\sin l[(1+f)^2s^2\epsilon_b^2 - \epsilon_\parallel^2].\nonumber
\end{eqnarray}
 From the definition of $f$ we see that these terms exclusively depend on the
measured distance $s' = (1+f)s$, so we can correct for proper-motion errors
before determining $f$.  Finally using the explicit form of $\vT$ from
Table~\ref{tab:mat} we obtain our correction terms:
 \begin{eqnarray}\label{eq:UWe}
\ex{e_Ue_x}&=&
-\fracj14\{\cos^2b\sin^22l\,[(s'^2(\sin^2b\epsilon_b^2
-\epsilon_l^2)+\cos^2b\epsilon_\parallel^2]\nonumber\\
&&\quad -\sin^22b\cos^2 l\,
[s'^2\epsilon_b^2 - \epsilon_\parallel^2])\}\\ 
\ex{e_We_y}
&=&\fracj14\sin^22b\{s'^2\epsilon_b^2 - \epsilon_\parallel^2\}.\nonumber
\end{eqnarray}

The left panels in \figref{fig:errdis} show $\ex{e_We_y}$ as a function of
the errors in the line-of-sight velocities (upper panel) and the errors in
proper motions (lower panel). (In the upper panels the proper-motion data are
error-free, while in the lower panels the line-of-sight velocities are
error-free.) All points are determined from realisations of a Monte-Carlo
sample of $450\,000$ disc stars and $50\,000$ halo stars sampled from the
model described by Table~\ref{tab:model}. The agreement between the analytic
formula and the Monte-Carlo results is perfect. On account of the large
distances of most of the stars, the proper-motion errors produce
substantially greater values of $\ex{e_We_y}$ than do the errors in $\vpa$
(which will be negligible for most present-day samples). The right panels of
\figref{fig:errdis} show the shifts in $f_1$ (red crosses) that arise from
the correlations plotted on the left. The uncorrected values of $f_1$ exhibit
a quadratic behaviour for small errors as can be expected from equation
(\ref{eq:UWe}), while for larger errors growth in the denominator on the
right of equation (\ref{eq:ffromW}) abates the growth in $|f_1|$. The blue
crosses show the values for $f_1$ obtained when we correct our estimate
according to equation (\ref{eq:UWe}). The green squares in the bottom right panel
depict the case when we vary $\sigma_{\mu_b}$ at a fixed line-of-sight
velocity error $\sigma_{\vpa} = 30 \kms$. This demonstrates that the error
effects can be added linearly and our formalism gives a perfect account of
them (purple circles).  We also checked that as predicted $\sigma_{\mu_l}$
does not affect our distance estimate when targeting $W$.

The largest uncertainty in the corrections given by equations (\ref{eq:UWe})
lies in the assessment of the measurement errors. The model data used above
include remote disc stars, which have errors that are larger than will often
be encountered in practice. So this test suggests that it should be possible
to correct for the effects of measurement errors in most samples.

\subsection{Rotation of the velocity ellipsoid}\label{sec:rotellipse}

Regardless of a star's location, we have been decomposing its velocity into
Cartesian components in the frame that is aligned with the Sun-centre line.
Since this frame is not aligned with the principal axes of the velocity
ellipsoid at the location of a distant star, we anticipate non-vanishing values
of $\ex{UV}$, etc., even in the absence of distance errors. We now address
this issue.

\begin{figure}
\epsfig{file=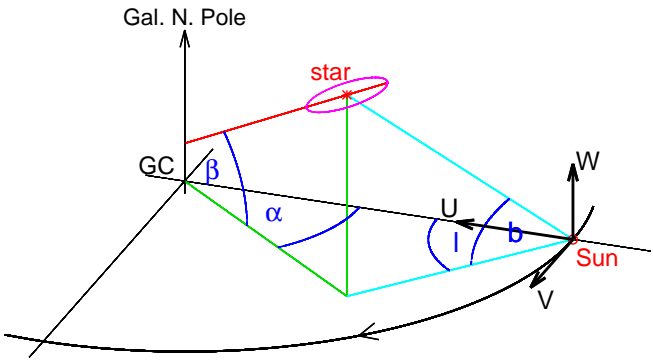,angle=-90,width=\hsize}
 \caption{The definition of Galactic coordinates, heliocentric velocities and
the angles $\alpha$ and $\beta$. GC signifies the Galactic Centre. The purple
ellipse depicts the direction of the radially oriented main axis of the
velocity ellipsoid (along $\Ug$), which defines $\beta$.}\label{fig:alpha}
\end{figure}

 Let the components of velocity of any star along the principal axes of its
local velocity ellipsoid be $(\Ug,\Vg,\Wg)$. Then with the angles $\alpha$
and $\beta$ defined as shown in \figref{fig:alpha}, the angle $\alpha$
between the projection onto the plane of the velocity ellipsoid's long axis
and the Sun-centre line is given by
 \begin{equation}\label{eq:defsalpha}
\alpha=\arctan\left({s\sin l\cos b\over R_0-s\cos l\cos b}\right),
\end{equation}
 and the heliocentric velocity components $(U,V,W)$ are given by
 \begin{equation}\label{eq:rotall}
\pmatrix{U + U_\odot \cr V + V_\odot \cr W +W_\odot}=
\vR(\alpha,\beta)\pmatrix{\Ug\cr\Vg\cr\Wg},
\end{equation}
 where
\begin{equation}\label{eq:defsR}
\vR(\alpha,\beta)\equiv\pmatrix{\cos\alpha\cos\beta & \sin\alpha & \cos\alpha\sin\beta \cr
-\sin\alpha\cos\beta & \cos\alpha & -\sin\alpha\sin\beta \cr
-\sin\beta & 0 & \cos\beta}.
\end{equation}

Both observation \citep{Siebert11} and theory \citep{BM11} suggest that
$\alpha$ and $\beta$ will take values close to the Galactocentric azimuth
$\phi$ and latitude $\frac12\pi-\theta$ of the location in question. In
our tests we will assume that these relations are exact.

\subsubsection{Correlations from mean streaming}\label{sec:rotation}

A major contribution to the velocity components $U$ and $V$ comes from the
azimuthal streaming of stars, which we take to have magnitude
$\vpbar(R,z)$.  This motion invalidates our assumption above that
$U=-U_\odot+\delta U$. Instead we now have $U=-U_\odot+\overline{U}+\delta
U$, where $\overline{U}$ is the $U$ component of the velocity field given by
$\vpbar$ at the star's location.  Specifically we have
 \begin{eqnarray} \label{eq:rot}
\overline{U}(s,l,b) &=& \vpbar \sin{\alpha},
\end{eqnarray}
where there is dependence on $(s,l,b)$ both through $\alpha$ and through the
(generally unknown) dependence of $\vpbar$ on $(R,z)$.  Unfortunately, both
$T_{UV}$ and $\overline{U}$ are odd functions of $l$, so correlations
contributed by distance errors can be interpreted as due to differential
rotation and vice versa. On account of this fact, $W$, to which $\vpbar$ does
not contribute, is a more useful target velocity than $U$. However, it is
nonetheless worthwhile to consider how $U$ can be targeted. 

 The first of equations (\ref{eq:otherUW}) now becomes
 \begin{eqnarray}\label{eq:rotUW}
U&=&-U_\odot+\vpbar\sin\alpha + fx'+(1+fT_{UU})\delta U.
\end{eqnarray}
To determine $f$ from these equations we
assume that 
\begin{equation}
\vpbar=\Theta g(R,z),
\end{equation}
 where $g(R,z)$ is a function that describes the way in which $\vpbar$ varies
with position and $\Theta\equiv \vpbar(R_0,0)$ is the local streaming
velocity of the population under study. In the simplest case we assume that
$g$ has no dependence on $R$, and we estimate its dependence on $z$ from the
data, using the current distance scale. Once $g$ has been chosen, and a
preliminary value for $\Theta$ adopted, we can determine the value of $x$ for
each star.  We primed $x$ in equation (\ref{eq:rotUW}) because $x$ contains the
mean motion in $U$, so we have to split off the rotation term:
 \begin{equation}
x' = x + T_{UU} \Theta g(R,z) \sin\alpha .
\end{equation}
 The distance error $f$ causes the measured $\alpha'$ to deviate from the
true value $\alpha$, but we can correct for this effect by  Taylor-expanding
$\alpha'(f)$:
\begin{equation}
\sin\alpha= \sin\alpha' - f\cos^3 \alpha' \frac{s' R_0 \sin l \cos b}
{(R_0 - s' \cos l \cos b)^2}+\hbox{O}(f^2).
\end{equation}
 Then
\begin{eqnarray}\label{eq:rotUWexp}
U &=& -U_\odot + \Theta \rho + fx + f\Theta k + (1+fT_{UU})\delta U,
\end{eqnarray}
where
\begin{eqnarray}
\rho&\equiv& g(R,z) \sin\alpha'\cr
k&\equiv&T_{UU} \rho + \cos^3 \alpha' \frac{s' R_0 \sin l \cos b}{(R_0 - s' \cos l \cos b)^2}. 
\end{eqnarray}
 Now we can proceed identically to the derivation of equation
(\ref{eq:ffromU}): we first subtract from equation (\ref{eq:rotUWexp}) its
expectation value to  obtain
\begin{eqnarray}
U - \ex{U} - \Theta (\rho - \ex{\rho}) &-& f(x-\ex{x}) \cr
 &-& f \Theta (k - \ex{k}) = (1+fT_{UU})\delta U,
\end{eqnarray}
 and then we multiply $x_i$ and $\rho_i$ and sum over our sample. Introducing
the abbreviation $s_{a,b} \equiv \Cov(a,b)$, this gives two equations for
the unknowns $f$ and $\Theta$:
\begin{eqnarray}
s_{Ux} - \Theta s_{\rho x} - f s_{xx} - \Theta f s_{kx} &=& f\ex{T_{UV}^2+T_{UW}^2}\sigma_U^2 \cr
s_{U\rho} - \Theta s_{\rho\rho} - f s_{\rho x} - \Theta f s_{k\rho} &=& 0 
\end{eqnarray}
 Inserting $\Theta$ from the second equation into the first and dropping all
terms of order $f^2$ we obtain our estimator
\begin{eqnarray}\label{eq:rotUfin}
f = \frac{s_{Ux}s_{\rho\rho} - s_{\rho x}s_{U\rho}}{s_{xx}s_{\rho\rho} - s_{\rho x}^2 + s_{kx}s_{U\rho} - s_{Ux}s_{k\rho} + t^2 \sigma_U^2 s_{\rho\rho}} 
\end{eqnarray}
 where $t^2 \equiv \ex{T_{UV}^2+T_{UW}^2}$. For a quick calculation the third
and fourth term in the denominator can be neglected as they are in general
small and only affect the slope.  Again we solve these equations iteratively,
at each iteration updating the distances and recalculating for each star $x$,
$\alpha$ and $g$.

\subsubsection{Correlations from random velocities}

In the heliocentric frame the random component $\delta U\equiv
U_0+U_\odot-\overline{U}$ is correlated with $\delta V\equiv
V_0+V_\odot-\overline{V}$ because the velocity ellipsoid at the star's
location is not aligned with that at the Sun's location, so the rotation
matrix $\vR(\alpha,\beta)$ of equation (\ref{eq:defsR}) is non-trivial.
Consequently, when we calculate $\ex{UT_{UV}V}$ in the course of evaluating
$\ex{Ux}$,
the correlation will be larger than the one we want by $\ex{\delta
UT_{UV}\delta V}$. We now determine the magnitude of this correlation so we
can subtract it from the correlations we obtain from the data prior to
determining $f$. Bearing in mind that $\ex{\delta\Ug\delta\Vg}=0$, we have
 \begin{eqnarray}
&&\ex{\delta UT_{UV}\delta V}=
\ex{(\delta\Ug\cos\alpha\cos\beta+\delta\Vg\sin\alpha + \delta \Wg\cos\alpha\sin\beta)\cr
&&\times T_{UV}(-\delta\Ug\sin\alpha\cos\beta+\delta\Vg\cos\alpha-\delta\Wg\sin\alpha\sin\beta)}\\
&&\quad=\fracj14\ex{\cos^2b\sin2l\sin2\alpha(\cos^2\beta
\sigma_U^2-\sigma_V^2 + \sin^2\beta\sigma_W^2)}\nonumber 
\end{eqnarray}
 Similar calculations yield the additional correlations
 \begin{eqnarray}\label{eq:shiftUW}
\ex{\delta UT_{UW}\delta W}& =&
 \fracj14\ex{\sin2\beta\sin2b\cos l\cos\alpha(\sigma_U^2-\sigma_W^2)}
\end{eqnarray}
 and
\begin{eqnarray}\label{eq:shiftVW}
\ex{\delta VT_{VW}\delta W} &=&
\fracj14\ex{\sin2b\sin l \sin\alpha \sin 2\beta (\sigma_W^2 - \sigma_U^2)}
\end{eqnarray}

\begin{figure}
\epsfig{file=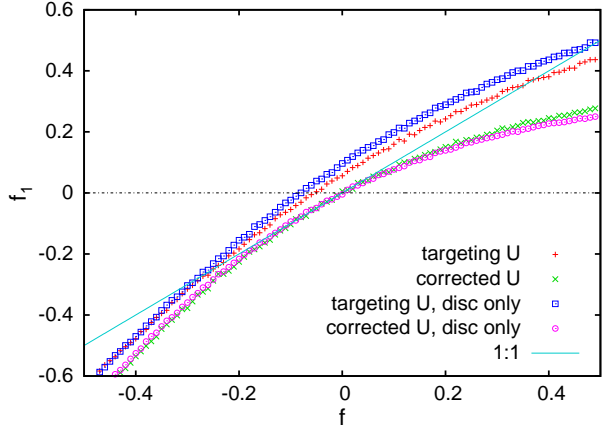,angle=-90,width=\hsize}
 \caption{The effect of including the corrections for rotation of the
 velocity ellipsoid. Two samples are used: one has just $500\,000$ disc
stars and the another has $450\,000$ disc plus $50\,000$ halo stars. Both samples are strongly affected
by rotation of the velocity ellipsoid, yet the correction successfully shifts
the points so they pass through the origin. As expected, the sample with
a halo contribution is less strongly affected.}\label{fig:rots}
\end{figure}

The red and blue points in \figref{fig:rots} show what happens if one ignores the
impact of azimuthal streaming and rotation of the velocity ellipsoids when
determining $f$ by plotting on the vertical axis the value of $f$ that is
recovered from equation (\ref{eq:ffromU}) against the input value of $f$. The
red points do not pass through the origin, so the estimated value of $f$ is
non-zero even when the distances are, in fact, correct. The green and blue
points show that when the formulae above are used to subtract the
contributions to the measured correlations from velocity-ellipsoid rotation,
the points pass through the origin as we require. The mock data used in these
tests consisted of $450\,000$ disc stars and $50\,000$ stars belonging to a
non-rotating halo in one case and a pure disc sample of $500000$ objects in the other case.  For one test case only the disc stars were used, while all the stars were used in the other case.

\subsection{Components with  extreme velocities}\label{sec:extreme}

Samples of halo stars generally have large mean $V$ velocities relative to
the Sun. So long as we are confident that the sample means of $U_0$ and
$W_0$ are far smaller than that of $V$, we can greatly simplify the analysis
of the
sample. While some samples of high-velocity stars may show a degree of radial
streaming on account of the Hercules star stream, the only indication of
streaming in the vertical direction is a very small correlation between $V$
and $W$ that was detected in the Hipparcos proper motions by \cite{Dehnen98},
and interpreted by him as the signature of the Galactic warp. We proceed
under the assumption that $\ex{U_0}=-U_\odot$ and $\ex{W_0}=-W_\odot$. 

At each point on the sky we imagine taking the sample mean of the third of
equations (\ref{eq:basicUVW}) to obtain
\begin{equation}\label{eq:givesx}
\ex{W}+(1+fT_{WW})W_\odot+fT_{WU}U_\odot= fT_{WV}\ex{V_0}
\end{equation}
 On the left we neglect terms of order $f$ and redetermine $f$ as the value
which gives the least-squares fit between the functions of sky coordinates
$\ex{W}+W_\odot$ and $T_{WV}\ex{V_0}\simeq T_{WV}\ex{V}$. The formal error in the
recovered value is
\begin{equation}
\epsilon_f={1\over\surd N}{\sigma_W\over\ex{V}\sigma_{T_{WV}}},
\end{equation}
 where $N$ is the number of bins on the sky.  For a typical sample
$\sigma_{T_{WV}} \sim 0.2$, and for halo stars we have $\ex{V} \sim 250 \kms$
and $\sigma_W \sim 100 \kms$, so $\epsilon_f \sim {2/\sqrt{N}}$, which gives an
error in $f$ $\epsilon_f\simeq6.3 \%$ for a sample of $1000$ objects.  We can
reduce this error by using the corresponding equation for $\ex{U}$ -- the
reduction is by a factor slightly smaller than $\surd2$ because
$\sigma_U>\sigma_W$.

If initially our distance scale is significantly in error, our first values
of $\ex{V}$ will be wrong. The magnitude of the problem is given by the first
term on the right of the second of equations (\ref{eq:basicUVW}):
\begin{equation}\label{eq:slowV}
\ex{V_0} \simeq \frac{\ex{V}}{1 + f\ex{T_{VV}}},
\end{equation}
 where the angle brackets around $T_{VV}$ imply the average over the surveyed
region of the sky. Eliminating $\ex{V_0}$ between  equations 
(\ref{eq:givesx}) and (\ref{eq:slowV}), we obtain
 \begin{equation}
\ex{W}+W_\odot\simeq{f\over1+f\ex{T_{VV}}}T_{WV}\ex{V}.
\end{equation}
 It is now straightforward to determine $f$ from the mean slope $x$ of the
correlation between $\ex{W}+W_\odot$ and $T_{WV}\ex{V}$:
\begin{equation}\label{eq:fextreme}
f={x\over1-x\ex{T_{VV}}}.
\end{equation}

This simple-minded approach to the determination of $f$ uses the available
information less efficiently than the technique described in Section
\ref{sec:sophisticated}, but it is a
good way of detecting a systematic distance error and its sign prior to
iteratively correcting the distance scale.  This is the approach that led
\cite{SAC} to suggest that the distances to low-metallicity stars in the
SEGUE dataset were being systematically overestimated.

\begin{figure}
\epsfig{file=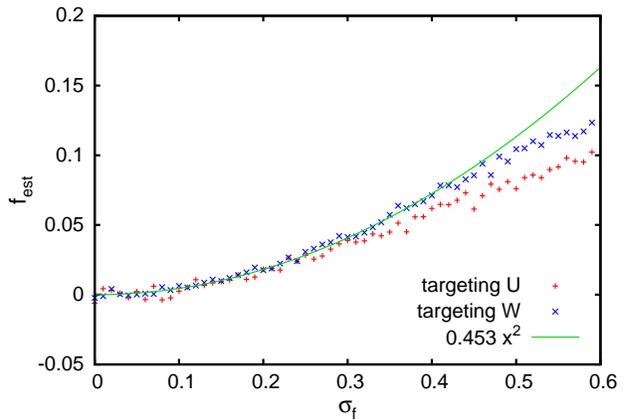,angle=-90,width=\hsize}
 \caption{The effect of an unbiased Gaussian distribution of distance errors
in a sample of $450\,000$ disc and $50\,000$ halo stars. For comparison we
plot the fits via $U$ and $W$ for the corrected fitting formulae.
At larger standard errors of the distances, the average distance estimate
diverges quadratically. For $W$ we show the fitting line $0.453 \sigma_f^2$
obtained for $0.0 < \sigma_f < 0.4$. Beyond this point the error loses Gaussianity because we have to cut the Gaussian distribution in order to avoid negative distances.}\label{fig:ferrb}
\end{figure}

\section{Scatter in the distance errors}\label{sec:scatter}

To this point we have assumed that the distances to all stars contain the
same fractional error, $f$. In reality any systematic offset will be combined
with random scatter, and we now consider whether in these circumstances the
factor $f$ that we recover from the whole population will equal the average
of the $f$ factors of the stars. In other words, does our procedure provide
an unbiased estimate of $f$?

\subsection{The bias in $f$}\label{sec:fbias}

In fact it is not hard to see from equation
(\ref{eq:ffromU}) that we must anticipate a tendency to overestimate the
mean value of $f$: stars with $f>0$ will be ascribed the largest velocities
and will thus tend to dominate the sums implicit in $\ex{Wy}$ and $\ex{y^2}$.
From the perspective that equations (\ref{eq:otherUW}) describe linear
relations between $U$ and $x$ or $W$ and $y$, stars with overestimated
distances will dominate the ends of the line and influence more
strongly our estimate of the line's slope $f$ than stars with under-estimated
distances, which will cluster near the middle of the line.

 \figref{fig:ferrb} shows this effect in samples of $450\,000$ disc and
$50\,000$ halo stars in which the input distances have errors $f$ that have
zero mean but the dispersion $\sigma_f$ that is given by the horizontal axis.
The vertical axis gives the recovered value of $f$. To both cases we apply the corrections
described in subsections \ref{sec:errors} and \ref{sec:rotellipse}.
The expected tendency for $f$ to be overestimated in the presence of
significant scatter in the input $f$ values manifests itself in the parabolic
shape of the curves formed by the corrected results. 

We can recover this behaviour analytically as follows. We assume
that the stars with a given fractional distance error $f'$ occur everywhere
on the sky, so we can form the sky-average $\ex{Wy}_{f'}$ over just this
group of stars. Defining
\begin{equation}
n^2 \equiv (T_{WV}^2 + T_{WU}^2)\sigma_W^2,
\end{equation}
 the inferred fractional distance error of the population
is
 \begin{eqnarray}
f&=&{\int\d f'\,P_f(f')\ex{Wy}_{f'}\over\int\d
f'\,P_f(f')\ex{y^2+n^2}_{f'}}\cr
&=&{\int\d f'\,P_f(f')f'\ex{y^2}_{f'}\over\int\d f'\,P_f(f')\ex{y^2+n^2}_{f'}},
\end{eqnarray}
 where $P_f(f')$ is the probability density function (pdf) of $f'$.
Now we we decompose $\ex{y^2}$ into the part $\ex{y^2}_\perp$ that derives
from tangential velocities and the part $\ex{y^2}_\parallel$ that derives from
line-of-sight velocities. Since the inferred tangential velocities scale like
$1+f$ we then have
\begin{equation}
f={\int\d f'\,P_f(f')f'[(1+f')^2\ex{y^2}_{f'\perp}+\ex{y^2}_{f'\parallel}]
\over
\int\d f'\,P_f(f')[(1+f')^2\ex{y^2+n^2}_{f'\perp}+\ex{y^2+n^2}_{f'\parallel}]}
\end{equation}
 Setting $P_f\propto\e^{-f^2/2\sigma_f^2}$ and neglecting the variation of
 $\ex{y^2}_{f'}$ with $f'$, this yields
 \begin{equation}\label{eq:fcorrect}
f\simeq{2\sigma_f^2\over1+\sigma_f^2+\ex{n^2}_\perp/\ex{y^2}_\perp + \ex{y^2+n^2}_\parallel/\ex{y^2}_\perp}.
\end{equation}
 The parabolic variation of the recovered value of $f$ with the width
$\sigma_f$ of the scatter in individual $f$-values is now manifest.

Actually, the assumption above of a Gaussian distribution of fractional
distance errors is not fully realistic. In fact,  the $P_f(f)$ has a
long tail at $f>0$. Stars in this tail will have seriously overestimated
tangential velocities, and  \cite{SAC} argue that as a consequence a halo
sample that in reality has no net rotation can be interpreted as consisting
of two populations, one of which is counter-rotating.

\subsection{The second moment of the error distribution}\label{sec:fsq}

We can obtain information about the breadth of the distribution of distance
errors in an approach largely similar to the classical statistic parallax: we compare the square of the speed $v$ with the squares of the
line-of-sight velocity $\vpa$.  For the measurement of the $i$th star, we
have
\begin{equation}
{v_{i}^2} ={v_{\parallel, i}^2 + F_i^2v_{\perp, i, 0}^2},
\end{equation}
 where $F_i\equiv1+f_i$.
Summing over the $N$ stars in the sample, we obtain
 \begin{eqnarray}\label{eq:vsqrat}
\frac{\ovs}{\ovpas} &=& 1+\frac{\sum{F_i^2 v_{\perp,
i0}^2}}{N\ovpas}\nonumber \\
&=& 1+  \frac{\ovfs \sum{v_{\perp, i0}^2}+\sum{(F_{i}^2 -
\ovfs)v_{\perp, i0}^2}}{N\ovpas}\\
&=&1 + \ovfs \frac{\overline{v_{\perp0}^2}}{\ovpas} 
+ {1\over\ovpas}\Cov\left(F^2, \overline{v_{\perp0}^2}\right).\nonumber
\end{eqnarray}
 If the distance errors are statistically independent of velocities, the
covariance vanishes. Further, if either the velocity distribution is
isotropic or the sample is uniformly distributed on the sky so $v_\perp$ 
and $\vpa$ sample equally all three principal axes of the velocity ellipsoid,
then $\overline{v_{\perp0}^2}=2\ovpas$ and equation (\ref{eq:vsqrat}) yields
\begin{equation}\label{eq:givesiso}
\ovfs=\fracj12\left({\ovs\over\ovpas}-1\right)\qquad\hbox{(isotropy)}.
\end{equation}
 If the velocity distribution is anisotropic and the sky coverage is
non-uniform, this formula will under-estimate $\ovfs$ when the sample points
towards the longest axis of the velocity ellipsoid and overestimate  it in
the contrary case. 

Classical statistical parallaxes are obtained under the assumption of
isotropic velocity dispersion, which is the circumstance in which equation
(\ref{eq:givesiso}) is most likely to hold, and clearly this equation is
closely related to the classical formula for a statistical parallax. The main
difference is that is yields the second rather than the first moment of $F$.

The covariance in equation (\ref{eq:vsqrat}) is non-vanishing when the
distance errors are not statistically independent of the velocities, for
example, because distances are more likely to be under-estimated when looking
into the plane than when looking to a Galactic pole.

In practice the scope for reliable application of equation (\ref{eq:vsqrat})
is limited since few samples are uniformly distributed on the sky and have
securely known values of the covariance term. 

A more effective way to determine the scatter in $f$  exploits the idea
introduced at the start of this Section that stars with overestimated
distances tend to have large values of $|x|$ and $|y|$, while stars with
under-estimated distances have small values of $|x|$ and $|y|$. Consequently,
if in equations (\ref{eq:ffromU}) or (\ref{eq:ffromW}) we restrict the sum to
stars with small (resp.~large) $x^2$ or $y^2$ we will probe the smallest
(resp.~largest) values of $f$ within the sample. By combining these estimates
of $f$ with the numbers of stars associated with each range of values of
$x^2$ or $y^2$, we can construct the probability distribution $P(f)$ of the
overall sample.

Whichever approach we adopt to determine the scatter in $f$, we should
take into account the errors in proper motions.  In the first approach they
increase $\overline{v_\perp^2}$ and $\overline{v^2}$, in the second approach
they push stars to large vales of $x^2$ and $y^2$ and thus affect the
distance estimator. Fortunately, in relatively nearby samples the impact of
errors in proper motions is limited.

\section{Implementation}\label{sec:implement}

In this section we explain which of the several formulae we have given for
the fractional distance error $f$ we recommend, and in what order they should
be used. Then we illustrate the procedure by applying it to a sample of stars
from the Sloan Extension for Galactic Understanding and Exploration \citep[SEGUE,][]{SEGUE}
and a sample from Data Release 8 of the Sloan Digital Sky Survey \citep[][]{SDSS3, SloanDR8}.

In any real data set there are likely to be stars with implausibly large
heliocentric velocities, and the first step should be to discard those stars.
We discard stars for with extreme galactocentric velocities, i.e. $|U|, |V| > 800\kms$ or $|W| > 400\kms$.
Then we bin the stars by some quantity of interest, such as surface gravity,
metallicity or value of $v_\phi$, and for each bin use equation
(\ref{eq:ffromW}) iteratively to determine $f$ for that group. The values of
$\ex{Wy}$ used in this equation are the raw values from the data minus the
corrections $\ex{e_We_y}$ from equations (\ref{eq:UWe}) and the corrections
for rotation of the velocity ellipsoid from equations (\ref{eq:shiftUW}) and
(\ref{eq:shiftVW}).  Once this stage in the analysis has been
completed, the distances of stars have been corrected for the most important
errors in the original data, and we may assume that any residual systematic
errors are small.

\begin{figure}
\epsfig{file=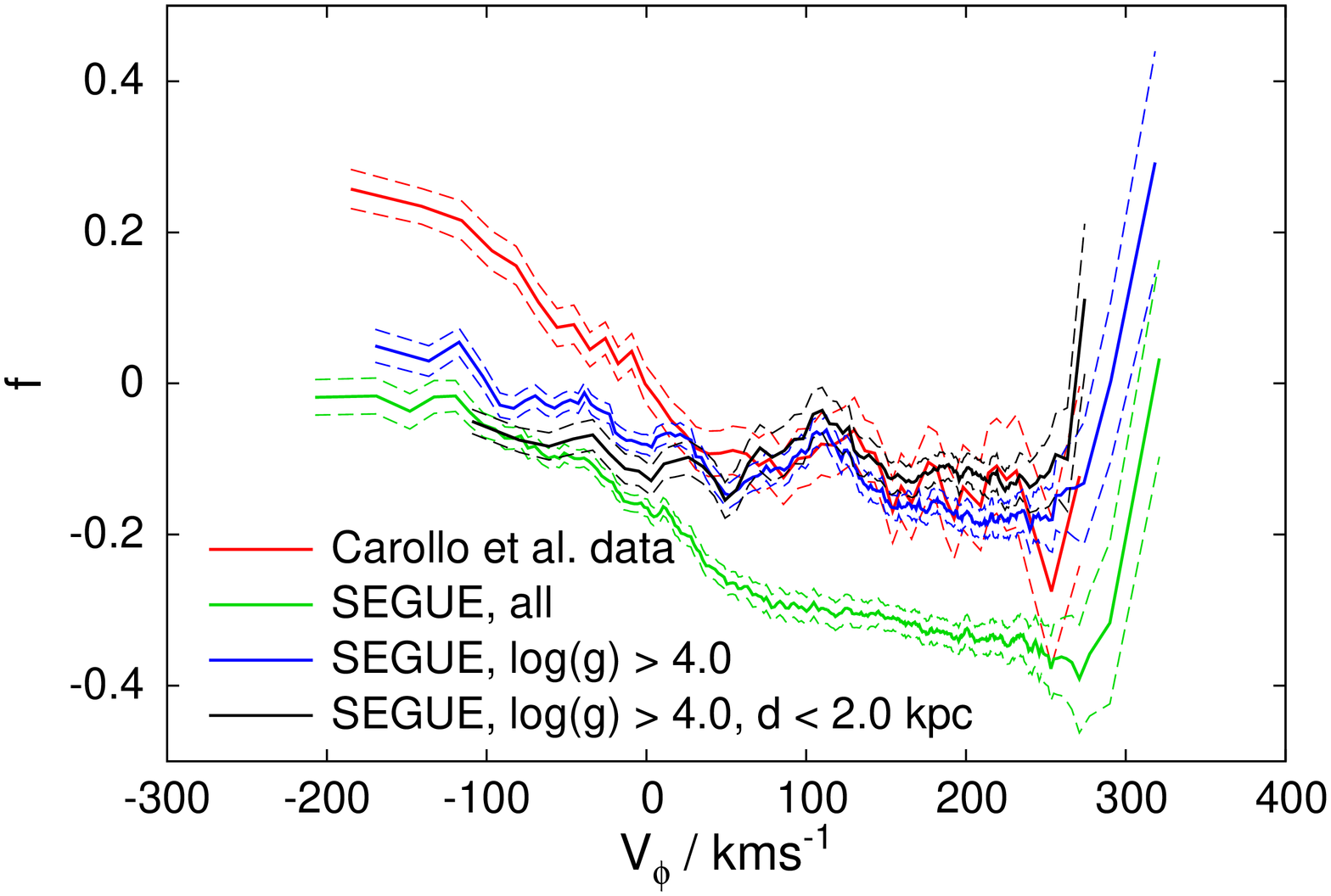,angle=-90,width=\hsize}
\epsfig{file=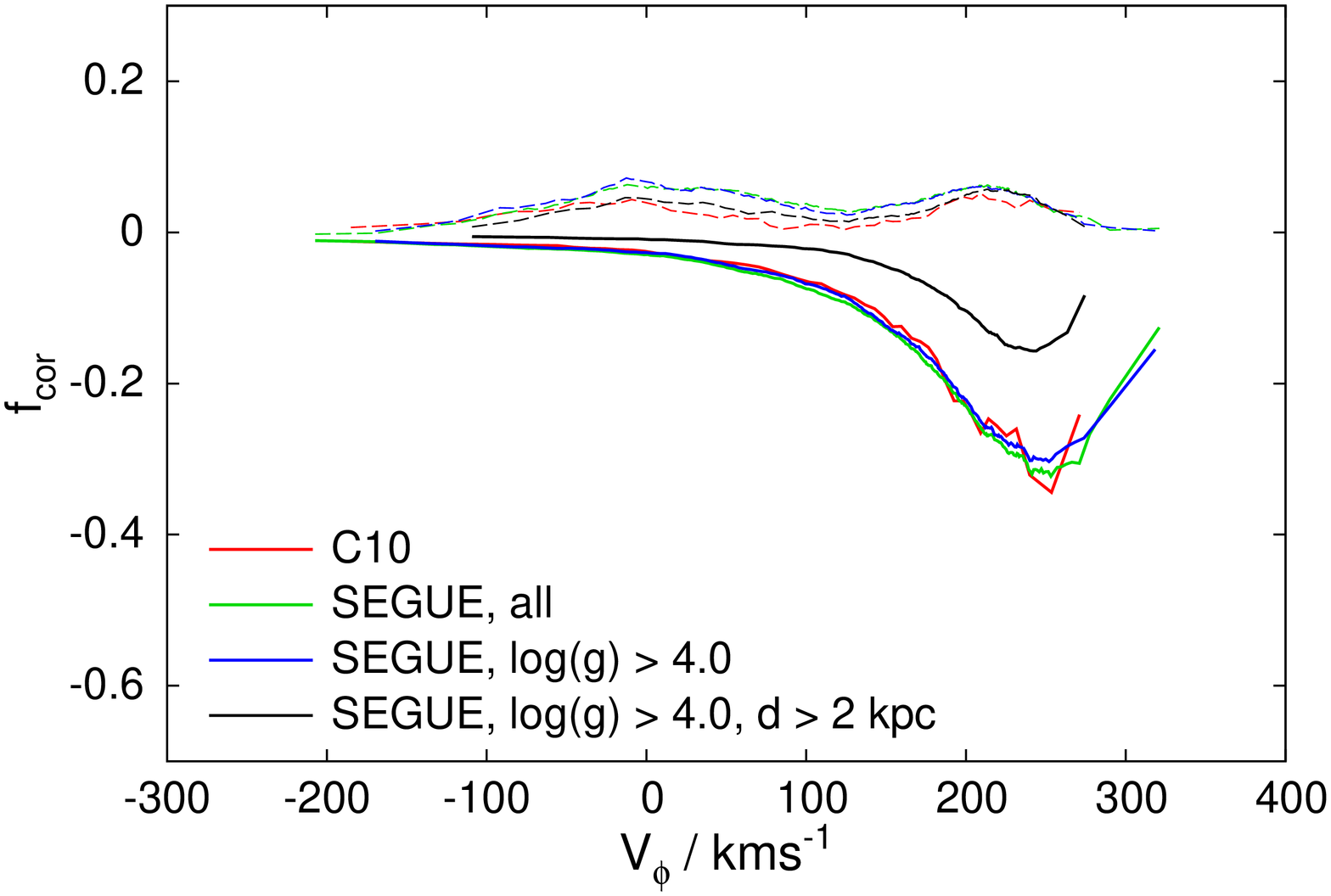,angle=-90,width=\hsize}
 \caption{Values of $f$ for the sample of $\sim20\,000$ main-sequence stars
 in the sample of Carollo et al.\ (2010) (red curve) and three different
selections on our SEGUE sample. A mask with a width of at least $1600$ stars
and at least $25 \kms$ width was moved over the sample in steps of $200$
stars. The dotted lines delimit the formal $1\sigma$ error bands associated
with these estimates. In the lower panel we plot the distance corrections
from proper motions (negative, solid lines) and the velocity ellipsoid turn
(positive, dashed lines) as defined in eq.~\ref{eq:shiftUW}.}\label{fig:fcar}
\end{figure}

\subsection{Used samples}

We will make use of two subsamples from the SEGUE survey. Our main sample
consists of a raw dataset of $224\,019$ stars from the eighths Sloan data
release \citep[DR8,][]{SloanDR8}. As we want the maximum number of stars we can get and not a
specific subset and do not fear metallicity biases in the sample, we use all
stars with clean photometry from target selection schemes that do not include
any direct kinematic (i.e. proper motion or line-of-sight velocity) bias. To
ensure decent quality of the used kinematics, we follow \cite{Munn04} in
requiring a match in the proper motion identifications ($match = 1$), a good
position determination $\sigma_{RA}, \sigma_{DEC} < 350 \,{\rm mas}$. To
ensure sensible stellar parameters, we require an average signal-to-noise
ratio larger than $10$. Further we require the formal errors on the proper
motions to be moderate: $\sigma_{\mu_b}, \sigma_{\mu_l} < 4 \,{\rm mas\,
yr}^{-1}$. We exclude any star that lacks a metallicity or a proper motion or
is flagged as having an unusual
 spectrum \citep[][]{Lee08a} unless the flag
indicates carbon enhancement. To eliminate a handful of objects with colours
far outside the normal calibration ranges we require $0 < (g-i)_0 < 1$. Only
stars that would be within $4 \kpc$ in the first guess distance determination
and pass our criteria for not being velocity outliers are used. When adopting
the \cite{Ivz08} (A7) main-sequence distance calibration a total of $119\,577$
stars pass these cuts. Velocities are derived as in \cite{SAC} with an
adopted solar galactocentric radius of $R_0 = 8 \kpc$, a circular speed of
$220 \kms$ and the solar motion relative to the local standard of rest from \cite{SBD}. For this work we make use of the dereddened Sloan photometric
colours provided in the catalogue.

The sample of \cite{Carollo10}, which comprises $\sim30\,000$ calibration stars
from SEGUE, constitutes our second sample. Its parameters derive from an
earlier version \citep[DR7,][]{SloanDR7} of the SEGUE parameter pipeline, but are consistent
with the new data release. While their sample is no more than a mere subset
of our larger sample, their sample suffers from distance
overestimates \citep[as shown by][]{SAC} whose re-detection illustrates the potential of the method presented here. 

\subsection{Mapping the samples in azimuthal velocity}

\figref{fig:fcar} shows the results of binning the main-sequence stars
of four subsamples of SEGUE stars by azimuthal velocity $v_\phi$ (with
the Sun at $v_\phi=232\kms$). The upper panel shows values of $f$, while the
lower panel shows the corrections used to obtain these $f$-values.

The full lines in the bottom panel of \figref{fig:fcar} show the corrections
to $f$ that are required to account for proper-motion errors -- the impact of
errors in $\vpa$ is negligible and not plotted. Proper-motion errors tend to
increase the recovered value of $f$, so they require a negative correction to
$f$. Their importance peaks around solar velocity because they
contribute a roughly constant term to $\Cov(W,y)$, while the typical
heliocentric velocities of stars, which provide our signal, shrink as
$v_\phi$ tends to the Sun's value both because of the diminishing offset in
the rotational component and because the velocity dispersion of disc stars
diminishes as $v_\phi$ approaches the circular speed.

The dashed lines in the bottom panel of \figref{fig:fcar} show the
corrections to $f$ that are required to account for the rotation of the
velocity ellipsoid. These curves have two peaks because there is a similar
competition between decreasing heliocentric velocities and decreasing size of
the velocity ellipsoid that drives the correction term.

The full red curve in the upper panel of \figref{fig:fcar} shows the values
of $f$ yielded by equation (\ref{eq:ffromW}) when distances from Carollo at
al.\ are used for their ``main-sequence'' stars; the dashed red curves show
the error bounds on $f$. We see that $f$ is significantly greater than
zero for $v_\phi< 0$, the region of retrograde rotation, implying the
presence of significant distance overestimates. At $v_\phi>0$, $f$ drops
slightly below zero. The full green curve shows the corresponding values of
$f$ for the full SEGUE sample when distances are obtained from the
\cite{Ivz08} (A7) main-sequence relation.  Since the samples are now much
larger, the formal error bounds are tighter than in the case of the Carollo
et al.\ sample. Now at $v_\phi>0$, $f$ is decidedly negative ($\sim-0.3$)
implying the presence of significant distance under-estimates.  The $f$-value
of a sample is an {\it average\/} distance correction, so a given value of
$f$ could imply that all stars have the corresponding distance mis-estimate,
or that a fraction of the stars have a larger mis-estimate while the bulk of
the stars have good distances.  The blue full curve in the upper panel of
\figref{fig:fcar} shows the values of $f$ obtained when the all-star sample
is restricted to dwarfs by imposing the restriction $\log g>4$: with this cut
the distances are under-estimated by only $\sim10$ per cent because the
gravity cut eliminates most sub-giants and giants from the sample. The black line shows the same ``dwarf'' star sample with the additional restriction for the primary distance estimate to be $d' \le 2 \kpc$. This cut removes mostly relatively blue stars that have a tendency to be on the blue side of the turn-off point. And as we can see from the black line in the lower panel the impact of proper motion errors is greatly reduced as these are proportional to the square of the estimated distance.

In light of this finding we conclude that the deep trough in the green curve for the
all-star sample arises because that sample is severely contaminated by
subgiants and giants.  We can probe the extent of the contamination by
dissecting a sample in velocity space because, as we saw in Section
\ref{sec:fbias}, stars with overestimated distances assemble at extreme
velocities, while stars with under-estimated distances are dragged towards
the solar motion.  This is why the curves of the two contaminated samples
(the Carollo et al.\ sample and the all-star sample) slope steeply downward
from left to right in the upper panel of \figref{fig:fcar}.  The slope of the
curve for the cleaner sample produced by the gravity cut is much smaller. We 
can even in analogy interpret the minor difference between the black and the 
blue curves: by the general inclination of the main sequence, the distance 
cut preferentially removes relatively bright blue stars from the sample that 
have a larger spread in estimated distances.

The sudden rise of $f$ at super-solar $v_\phi$ has a different cause.  These
stars are few in number and have small heliocentric velocities, and, as the
lower panel of \figref{fig:fcar} shows, their $f$-values are strongly
affected by assumed proper-motion errors.  It is likely that our probably
false assumption of a constant proper-motion error has biased the $f$-values
for these stars. By contrast, the $f$-values of stars with $v_\phi$
substantially smaller than the solar value are insensitive to the handling of
proper-motion errors.

While valuable insights can be obtained by examining $f$ as a function of
velocity, a word of caution about such dissection is in order.  No cut or
selection should directly affect the target variable (here $W$), and cuts on
the explaining variable can introduce artifacts that should be explored with
mock data. For example cutting in the heliocentric $V$ velocity instead of
$v_\phi$ introduces a velocity-dependent error in $f$ of order $\sim 5$ per
cent.  In this case the bias arises from the rotation of the velocity
ellipsoid, which from selection in $V$ creates a bias in $U$, which in turn evokes
biases in $W$ through the vertical tilt of the ellipsoid.

\begin{figure*}
\epsfig{file=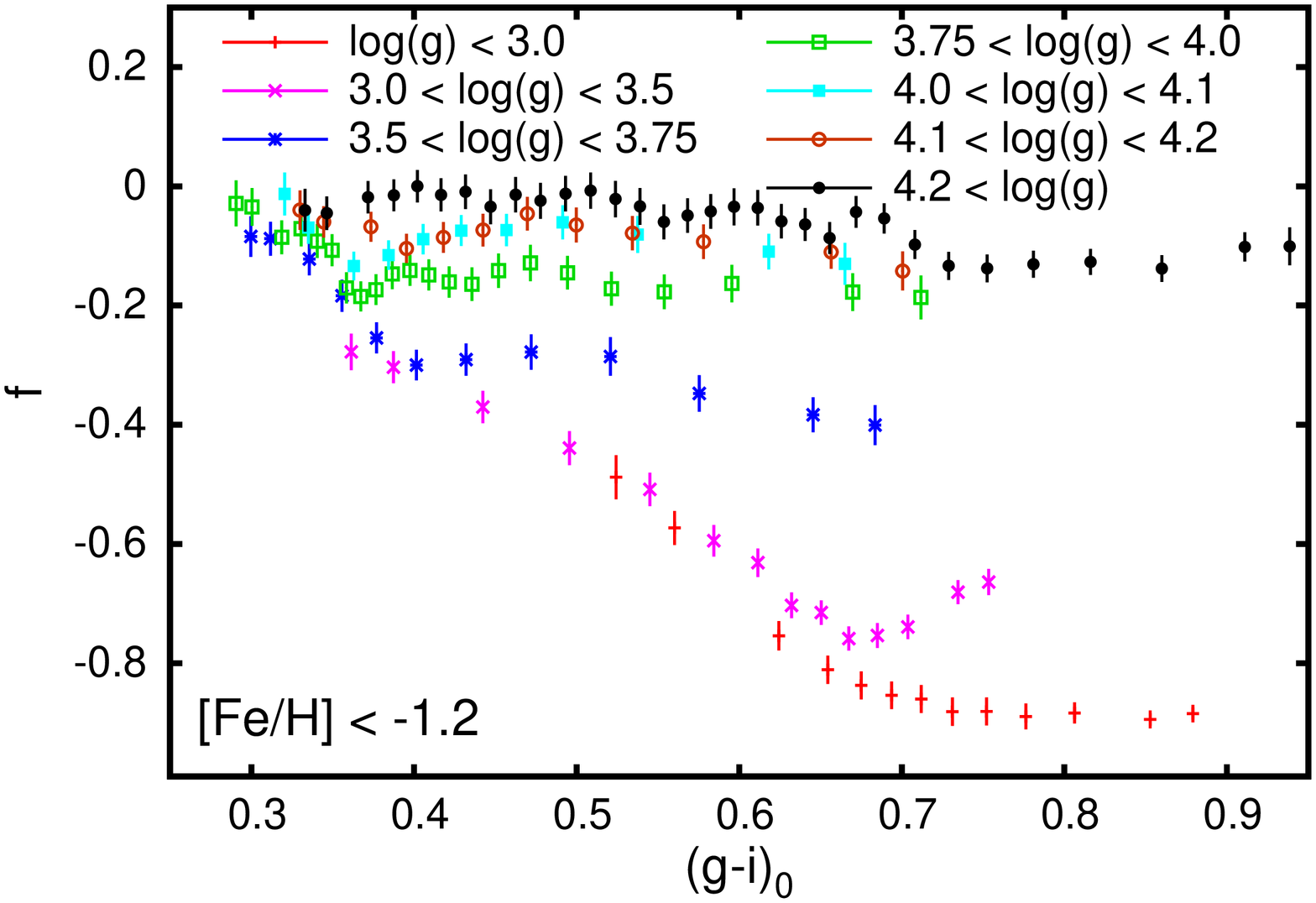,angle=-90,width=0.49\hsize}
\epsfig{file=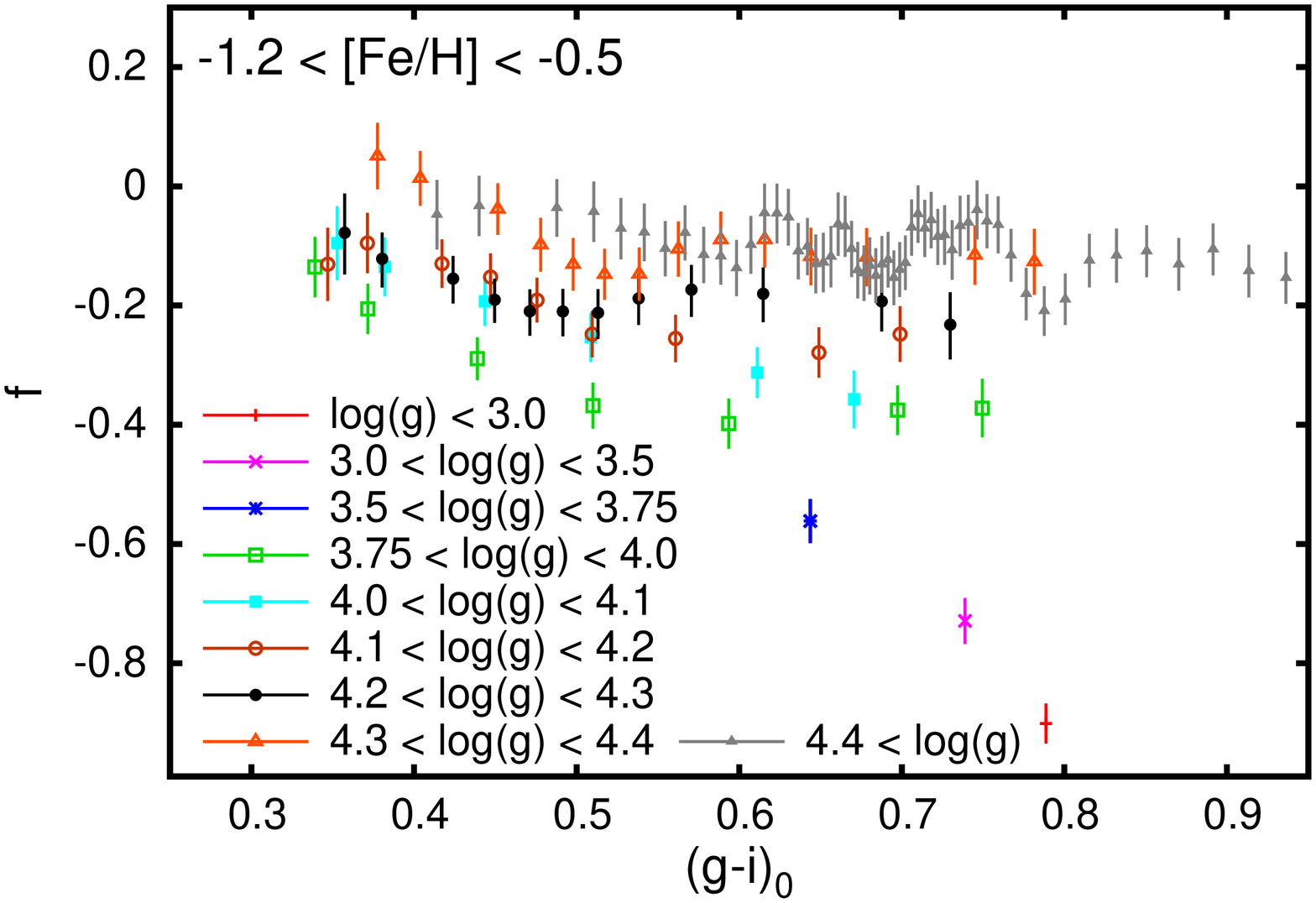,angle=-90,width=0.49\hsize}
\epsfig{file=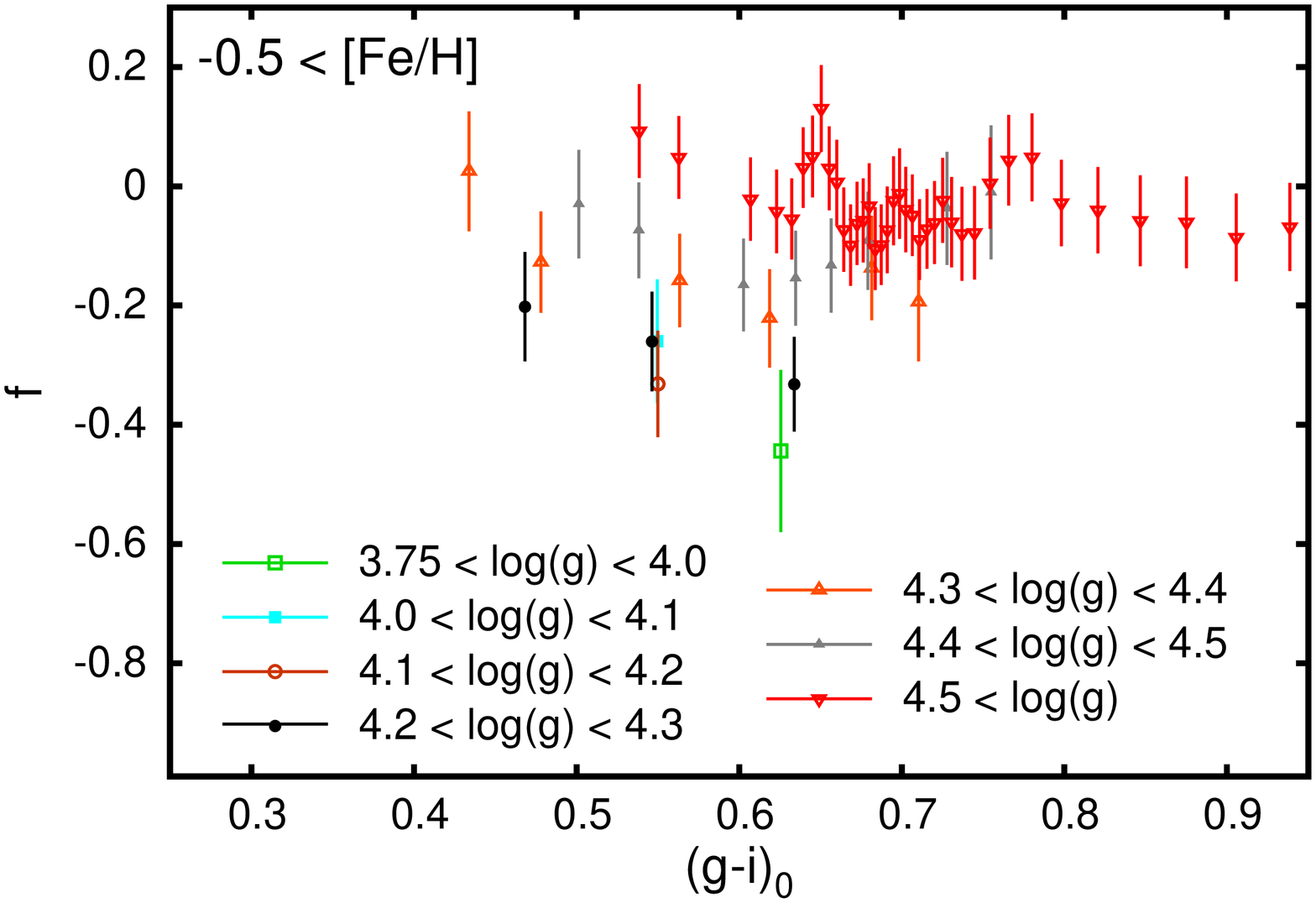,angle=-90,width=0.49\hsize}
\epsfig{file=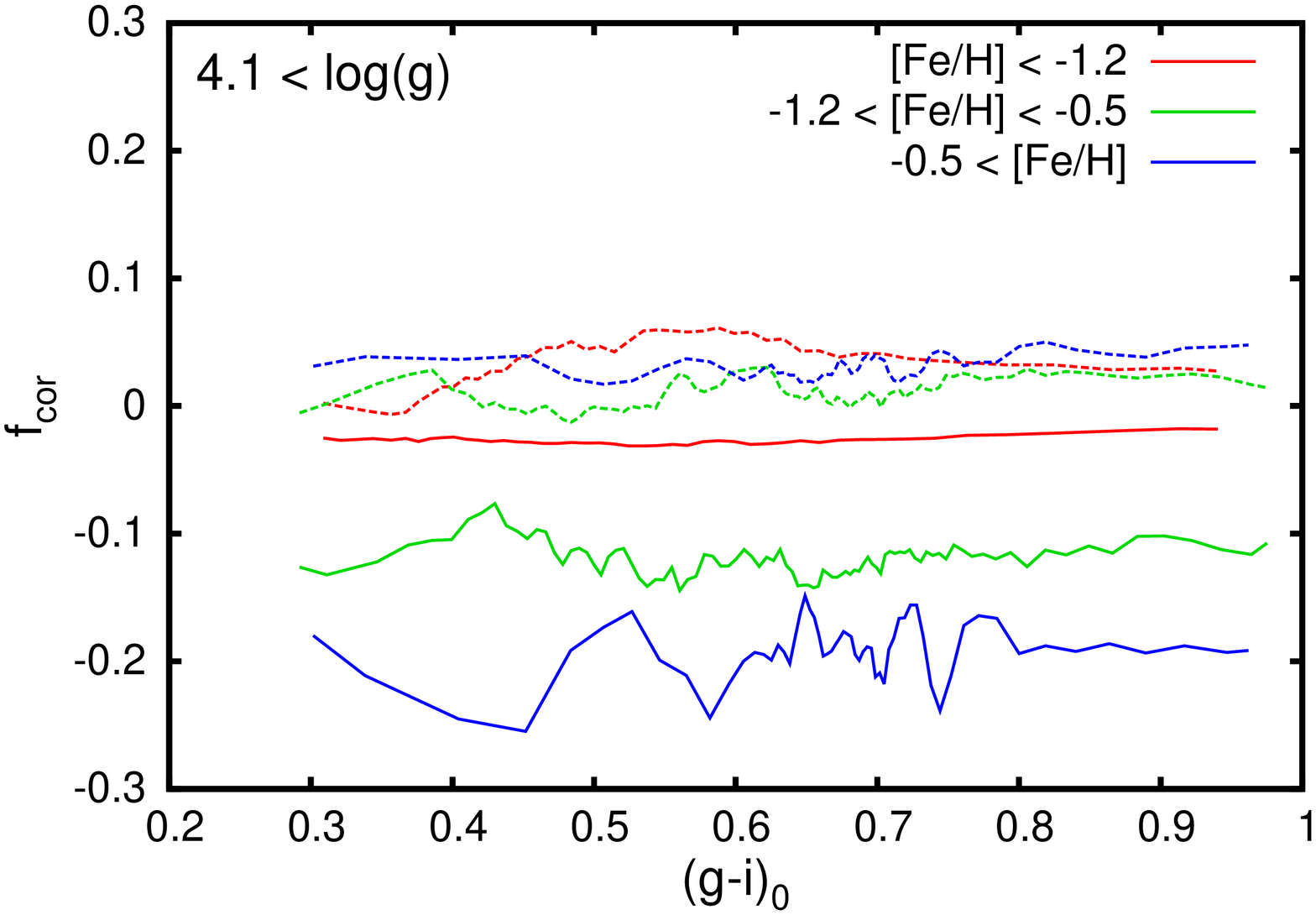,angle=-90,width=0.49\hsize}
 \caption{An evaluation of the performance of gravities in SEGUE DR8 and the
performance of the Ivezi{\'c} et al.(2008, A7) calibration against metallicity and
colour. Each of the first three panels shows results for stars in a restricted
range of metallicity. Each metallicity group was then divided by surface
gravity and $g-i$ colour and $f$ determined for that group from equation
(\ref{eq:ffromW}). We move a $1200$ stars wide mask over the sample in steps
of $400$ objects, so that every third data point is fully independent. Error
bars give the formal error plus a $30\%$ error on the systematic corrections.
The bottom right panel shows the corrections made to the $f$-values of
high-gravity stars of various metallicities for proper motions (solid
lines) and the turn of the velocity ellipsoid (dashed lines).
\label{fig:colseg} }
\end{figure*}

\subsection{Dissecting the main sample in gravity}\label{sec:Ivezic}

By partitioning a sample in gravity we can explore the extent to which a
sample contains stars at different evolutionary stages since they should fall
into different bins in gravity. In the following we use only the A7
calibration of \cite{Ivz08}.  

The first three panels of \figref{fig:colseg} show results obtained by
splitting the $\sim120\,000$ stars into three ranges of metallicity, with
boundaries at $\hbox{[Fe/H]}=-1.2$ and $-0.5$ and then within each
metallicity group splitting the stars in $\log g$, and finally binning them
in colour. Each colour bin has $\geq 1200$ stars (the ones at the edges carry 
$\geq 800$ and $\geq 400$ objects), and from one bin to the
next $400$ stars are dropped, so every third data point is independent. Points
are plotted at the average colour of the stars in the bin. Most giant stars
in this sample have low measured metallicities so the low-gravity bins are
only well-populated for the most metal-poor stars.  The error bars indicate
the formal errors on $f$ plus an error of $30\%$ in the corrections to $f$
for proper-motion errors and rotation of the velocity ellipsoid.

Since the distances employed assume that every star is on the main sequence,
giants have severe distance under-estimates (negative $f$). In the top two
panels of \figref{fig:colseg} one can assess the colour at which stars move
up from the subgiant branch to the giant branch -- the precision with which
this colour can be determined is increased if the sample is not divided by
gravity or metallicity. The distance under-estimates indicated by
\figref{fig:colseg} are similar to those we would expect a priori, but the
agreement is imperfect because the giants in this sample are very remote, so
proper-motion errors have a big impact on kinematically determined distances.

In the literature SEGUE stars with gravities within $3.0 < \log g < 3.5$ are
considered subgiants \citep[see e.g.][]{Carollo10}. Stars with $3.5 < \log g
< 4.0$ were classified as turn-off stars until it was shown by \cite{SAC}
that this practice sorts stars into unphysical positions in the
colour-gravity plane (at the relevant low metallicities, the turn-off region
should end bluewards of $(g-i)_0 < 0.4$).  More recent studies \citep[][]{Beers11,Carollo11} classify the
stars with $3.5 < \log g < 3.75$ as subgiants and the higher-gravity objects
as main-sequence stars. However, the purple and blue points in the upper two
panels of \figref{fig:colseg} show that it cannot be the case that all stars
with $\log g<3.75$ are subgiants, both because at $(g-i)_0>0.4$ the
$f$-values of the stars with $3.5<\log g<3.75$ are significantly less negative than those
of stars with $\log g<3.5$, and because the $f$-values of the high-gravity
sub-sample are no smaller than $\sim-0.3$. This corresponds to their being
more luminous than main-sequence stars of the same colour by less than a
magnitude, whereas, depending on metallicity, already at $(g-i)_0 \sim 0.4$
subgiants should be more luminous than main-sequence stars by more than $1.5$
magnitudes.  We conclude that no reliable selection for subgiants is
feasible with the current gravities: in general there is a contamination by
dwarf stars (with the well-known effects of distance overestimates mimicking
kinematically hot retrograde populations) and at least on the red side we
have to expect some contamination by giants.

The main-sequence relation appears to describe relatively well the distances
of stars with measured $\log g>3.75$. Yet, especially in the top right panel of
\figref{fig:colseg} we see that for all metallicity subsamples, $f$ tends to
increase bluewards. This phenomenon arises because the colour-luminosity
relation we have used is inclined relative to the theoretical zero-age main
sequence and assigns quite high luminosities and consequently large distances
to blue stars relative to their red counterparts. 

\figref{fig:colseg} enables us to choose the lower limit on gravity that will
most effectively minimise contamination of the final sample by stars that are
not dwarfs. This limit appears to rise from about $\log g \sim 4.1$ at the
lowest metallicities to $\log g \sim 4.4$ at the highest metallicities. Some part 
of the trend may also be connected to the redward shift of the turn-off with metallicity. 
However, this conclusion should not be blindly transferred to catalogues other than
DR8 because in this parameter derivation measured gravity is likely correlated with
metallicity, so we may to some extent see mapping errors in assumed luminosity
that arise from errors in metallicity. \figref{fig:colseg} also enables us to
detect the redward shift with increasing [Fe/H] in the turn-off colour
as the colour at which the dark-blue points of lower-gravity stars become
clearly separated from the black points of dwarf stars. Also, blueward of the turn-off we
expect an increased spread in values of $f$ within the highest-gravity bins, as the SEGUE
stellar parameter pipeline retains some residual information on how high above the main 
sequence a star is placed in gravity.

The bottom-right panel of \figref{fig:colseg} shows the corrections to $f$ required
by proper-motion errors (solid lines) and rotation of the velocity ellipsoid
(dashed lines).  The impact of proper-motion errors on the most metal-weak
stars is small because these stars are in the halo and have large
heliocentric velocities. Rotation of the velocity ellipsoid has similar
impact on stars of all metallicities because these stars are distributed
through broadly the same volume and a higher velocity dispersion both
inflates the correction and the signal in $f$.  For this plot velocity errors
were calculated by measuring the dispersions in each subsample and then
assuming constant velocity dispersions in the lowest metallicity bin and in
the other metallicity bins an increase of the dispersion by $15$ per cent for
each kiloparsec in $|z|$ and assuming that
$\ex{U^2}^{1/2}\propto\exp(-R/R_\sigma)$ with $R_{\sigma} = 7.5 \kpc$. This
correction term is small and minor changes in how it is
derived will not alter our results.

\section{Conclusions}\label{sec:conclude}

Systematic distance errors give rise to correlations between the measured
components $(U,V,W)$ of heliocentric velocities. Similar correlations arise
from three other sources: (i) measurement errors in the proper motions, (ii)
Galactic streaming motions and (iii) dependence of the orientation of the
local velocity ellipsoid on position in the Galaxy. However, each of these
sources of correlation between $(U,V,W)$ has a different and known pattern of
variation over the sky, so provided the data come from a wide-area survey, we
can disentangle their effects. We have described an iterative procedure by
which the distances to stars are rescaled until correlations between
$(U,V,W)$ are fully accounted for by effects (i) -- (iii) above, and the contribution from systematic distance errors vanishes.

The procedure works best when the group of stars under study has a large net
motion with respect to the Sun. This net motion is often dominated by
azimuthal streaming, since the Sun has more angular momentum than a
circular orbit, while nearby thick-disc and halo stars tend to have
substantially less angular momentum. When azimuthal streaming is dominant,
the simpler formulae of Section \ref{sec:extreme} apply. For stars in the
thin disc that have similar angular momentum to that of the Sun the other
velocity components usually still carry sufficient information to assess distances.

In principle we can determine distance errors by using either $U$, $V$ or $W$
as a ``target'' variable, with the ``explaining'' variable being composed of
the other components of velocity and the sky coordinates. In practice $V$
should not be used as a target variable as the systematic variation of $V$
velocities with position in the Galaxy would invoke spurious correlations
with the angle terms connecting it to the explaining velocity components.
$W$ is the target variable of choice both because it has the smallest
velocity dispersion and because it is least affected by the complexities of
differential rotation. $U$ is mainly useful as a target variable for its
ability to determine the mean rotation rate of a population once the distance
scale has been corrected by exploiting $W$. We will discuss an application to
this rotation term in a forthcoming paper.

There are some restrictions on the applicability of the method that
should borne in mind when using it. The proper motions need to be unbiased
and their errors should have finite and approximately known variances.  These
conditions seem to be satisfied by data from the SDSS \citep{Dong11}. If the
sample is non-local we need to estimate the extent of rotation of the
velocity ellipsoid within the sampled region. Such an estimate can be
obtained from the sample itself, but with some residual uncertainty arising
from proper-motion errors that particularly affect remote stars. 

Streams and
a warp will induce unwanted correlations but the likelihood of these giving
rise to an erroneous distance scale is small for several reasons. 
First, for a
stream or warp to undermine the method, the correlations it introduces must
vary on the sky in a similar way to the correlations associated with distance
errors. Consequently, the impact of a stream or warp is likely to be
suppressed given sufficient sky coverage. Second, a warp could be accounted
for in much the same way we have accounted for Galactic rotation. Third,
the footprints of streams or a warp will show up in conflicting values
for $f$ obtained from the two possible target velocities, $W$ and $U$.
Finally, a stream or warp would induce identical correlations in the
velocities of stars in the broad colour and gravity range that made up the
physical feature, whereas distance mis-estimates will usually vary with
spectral type.

In Section \ref{sec:fbias} we showed that the estimators given in Section
\ref{sec:af} are mildly biased in the sense that when there is a scatter in
the distribution of distance errors, the estimated value of $f$ will be
larger than it should be by an expression quadratic in the width of the
distribution. Equation (\ref{eq:fcorrect})  can be used to correct for this
effect. As we discussed in Section \ref{sec:fsq}, the method can be extended
to probe the full probability distribution of $f$ values rather than just
determining the mean value of $f$. Details of this extension will be given in a
later paper. 

In Section \ref{sec:implement} we applied the method to samples of stars from 
the SEGUE survey. We concluded that the distances to stars used by \cite{Carollo10} 
are on average significantly overestimated among stars deemed to be counter-rotating, 
and tend to be under-estimated by $\sim10$ per cent near solar velocity. This is also a 
nice example on how a spread in the distance errors within a sample can be directly seen 
by eye, when we dissect the sample in velocity: the distance overestimates assemble at
velocities remote from the solar value (i.e. mostly the retrograde tail of the halo
velocity distribution), while in our all-star sample, which is contaminated by 
numerous giants, the giants are dragged towards the solar motion, so a trough forms 
in a plot of the correction factor $f$ versus azimuthal velocity. 
We also warn against mistaking the derived values for a direct estimate 
of absolute magnitudes. Apart from contaminations the method corrects for the mean 
reddening error and the presence of unresolved binaries. So in general it can be expected
to give slightly larger mean distances than appropriate for single stars.

In Section \ref{sec:Ivezic} we demonstrated the use of the method to
assess the reliability of the  Ivezic (2008) A7 distance scale for dwarfs and
to assess the degree of contamination by non-dwarfs that arises as the lower
limit on $\log g$ for entering the sample is varied.
For low contamination  the lower limit on $\log g$ should increase 
with metallicity. We conclude that using only the DR8 gravities it is not
possible to achieve a satisfying selection of
subgiants. The level of
contamination by dwarf stars becomes large once the upper limit on $\log g$
exceeds $\sim3.5$. Since any dwarf that is misidentified as a subgiant has a
seriously overestimated distance, studies of stellar kinematics that are
based on DR8 gravities should rigorously exclude subgiants.

We are currently applying the method to recent distances to stars in the RAVE
survey \citep{Zwitter11,Burnett11}. A wide variety of applications to this
method will follow as it offers a standard tool to identify groups of stars
with problematic parameters, to check the reliability of selection schemes and
distance assignments and finally to correct for any biases in these
distances, e.g. by deviations in reddening with distance from the Sun. 

Our study has also illuminated the kinematic patterns that distance errors
can generate. These are not limited to the production of spurious
counter-rotating components, but include tilts of the velocity ellipsoids,
and by allowing rotational velocity to masquerade as motion in either the
radial or vertical direction, can extend to patterns of mean motion that, in a sample
that is anisotropically distributed on the sky, can imply a wrong motion of
the local standard of rest.

\section*{Acknowledgements}
R.S. acknowledges financial and material support from
Max-Planck-Gesellschaft. We thank Michael Aumer for the kind provision of his galaxy
models and Ruobing Dong for helpful discussions on SDSS proper motions. We thank M. Williams 
for careful reading and comments.

\label{lastpage}

\end{document}